\documentclass[11pt,journal,onecolumn,draftcls]{IEEEtran}
\usepackage [dvips]{graphicx}
\usepackage{subfig}
\usepackage{psfrag}
\usepackage{multirow}
\usepackage[cmex10]{amsmath}
\usepackage{amssymb}
\usepackage{psfrag}
\usepackage{bbm}
\DeclareSymbolFont{bbold}{U}{bbold}{m}{n}
\DeclareSymbolFontAlphabet{\mathbbold}{bbold}

\newtheorem{theorem}{Theorem}
\newtheorem{lemma}{Lemma}
\newtheorem{corollary}{Corollary}
\newtheorem{definition}{Definition}
\newtheorem{proposition}{Proposition}
\newtheorem{remark}{Remark}

\begin{document}

%%%%%%%%%%%%%%%%%%%%%%%%%%%%%%%%%%%%%%%%%%%%%%%%%%%%%______Title______%%%%%%%%%%%%%%%%%%%%%%%%%%%%%%%%%%%%%%%%%%%%%%%%%%%%%%%
\title{Capacity and Rate Regions of A Class of Broadcast Interference Channels}
\author{Yuanpeng~Liu,
        Elza~Erkip,~\IEEEmembership{Fellow,~IEEE}%
\thanks{The material in this paper was presented in part at the 2012 IEEE Int. Sym. Inf. Theory and 2013 IEEE Int. Sym. Inf. Theory.}%
\thanks{Y. Liu was with the Department of Electrical and Computer Engineering, Polytechnic School of Engineering of New York University, Brooklyn, NY, USA (e-mail: liuyp1984@gmail.com).}%
\thanks{E. Erkip is with the Department of Electrical and Computer Engineering, Polytechnic School of Engineering of New York University, Brooklyn, NY, USA (e-mail: elza@nyu.edu).}}

\maketitle

%%%%%%%%%%%%%%%%%%%%%%%%%%%%%%%%%%%%%%%%%%%%%%%%____Abstract_____%%%%%%%%%%%%%%%%%%%%%%%%%%%%%%%%%%%%%%%%%%%%%%%%%%%
\begin{abstract}
In this paper, a class of broadcast interference channels (BIC) is investigated, where one of the two broadcast receivers is subject to interference coming from a point-to-point transmission. For a general discrete memoryless broadcast interference channel (DM-BIC), an achievable scheme based on message splitting, superposition and binning is proposed and a concise representation of the corresponding achievable rate region $\mathcal{R}$ is obtained. Two partial-order broadcast conditions {\em interference-oblivious less noisy} and {\em interference-cognizant less noisy} are defined, thereby extending the usual less noisy condition for a regular broadcast channel by taking interference into account. Under these conditions, a reduced form of $\mathcal{R}$ is shown to be equivalent to a rate region based on a simpler scheme, where the broadcast transmitter uses only superposition. Furthermore, if interference is strong for the interference-oblivious less noisy DM-BIC, the capacity region is given by the aforementioned two equivalent rate regions. For the interference-cognizant less noisy DM-BIC, it is argued that the strong but not very strong interference condition does not exist and in this case, the capacity region for the very strong interference is obtained. For a Gaussian broadcast interference channel (GBIC), channel parameters are categorized into three regimes. For the first two regimes, which are closely related to the two partial-order broadcast conditions, achievable rate regions are derived by specializing the corresponding achievable schemes of DM-BICs with Gaussian input distributions. The entropy power inequality (EPI) based outer bounds are obtained by combining bounding techniques for a Gaussian broadcast channel (GBC) and a Gaussian interference channel (GIC). These inner and outer bounds lead to either exact or approximate characterizations of capacity regions and sum capacity under various conditions. For the remaining complementing regime, inner and outer bounds are also provided.
\end{abstract}

%%%%%%%%%%%%%%%%%%%%%%%%%%%%%%%%%%%%%%%%%%%%____Introduction_____%%%%%%%%%%%%%%%%%%%%%%%%%%%%%%%%%%%%%%%%%%%%%%%
\section{Introduction}
Broadcast channel (BC) and interference channel (IC) are two important classes of multi-user channels that have drawn considerable research attention in the past few decades, mostly due to their simplicity as fundamental building blocks and their close relevance to practical communication networks. While complete capacity characterizations are not available, there have been significant advances on these topics in the information theory literature. Notably the best general achievable schemes for the two channels are respectively given by Marton \cite{Marton} and Han-Kobayashi \cite{Han}, which are capacity achieving for some subclass channels or under various conditions, such as the ones in \cite{Cover BC}-\cite{Etkin}.

\begin{figure}
  \centering
  \includegraphics[width=50mm]{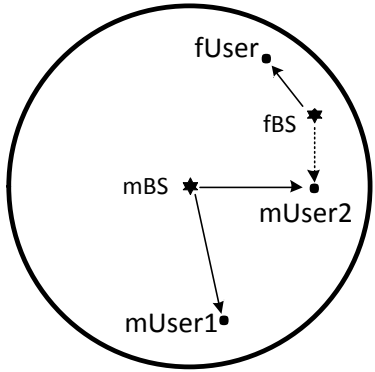}
  \caption{A macro base station (mBS) communicates with two macro users, one of which is interfered by the transmission between a femto base station (fBS) and a femto user}
  \label{fig:cell}
\end{figure}

Motivated by a recent interest in a heterogeneous cellular network design paradigm \cite{Andrews}, we explore a multi-user channel that combines the broadcasting and interference components, i.e. broadcast interference channel. This channel models communication scenarios that can be easily found in heterogeneous cellular networks. For example, in a macro-femto setting in Fig. \ref{fig:cell}, the BIC describes a scenario, where a multi-user macro cell interferes with a single-user femto cell. A fully-connected BIC consists of interference both from the macro base station (BS) to the single femto user and from the femto BS to all the macro users. In this paper, we focus on a simplified subclass of BICs with the following assumptions: 1. there are two macro users; 2. only one of them is interfered by the femto transmission; 3. the macro BS does not interfere the femto user. Item 1 represents the the simplest nontrivial broadcast configuration. The justification for item 2 is that, with the fair assumption of a uniform placement of macro users in the cell, the chance that both of the users are simultaneously close to the femto BS is small. Item 3 is also not unreasonable because signal-to-interference-noise ratio in the femto cell is likely to be high enough such that treating interference as noise is close to optimal. Hence any sensible scheme that works with the simplified BIC can be easily adapted to the case where item 3 does not hold without incurring excessive loss of performance. At last, we believe that a fundamental understanding of this simplified channel is crucial for characterizing the trade-offs in a more complex heterogeneous network.

Variations of BICs have been previously studied by Shang and Poor in \cite{Shang I}, for a different interference pattern where interference is generated by the broadcast transmitter, and in \cite{Shang II}, for the GBIC with a similar interference pattern as in this paper where both of the broadcast receivers are subject to interference. Even though the channel studied in this paper has a more restrictive interference pattern, we derive tighter outer bounds for the Gaussian channel, we also address the more general discrete memoryless channel and we provide more general classes of common strategies as well as capacity characterizations under various conditions.

In this paper, we first focus on a DM-BIC. Specifically we propose a general achievable scheme that is a natural combination of Marton and Han-Kobayashi schemes. In particular, the broadcast transmitter employs message splitting, superposition and binning, the interfering transmitter employs message splitting and superposition, and the receivers perform joint-typicality decoding. A concise representation of the corresponding achievable rate region $\mathcal{R}$ is obtained. We then extend the less noisy condition for a broadcast channel by taking interference into account and define two partial-order broadcast conditions {\em interference-oblivious less noisy} and {\em interference-cognizant less noisy}. Under these conditions, we consider a simplified scheme where binning is removed, resulting in reduced forms of $\mathcal{R}$, denoted by $\mathcal{R}_i$, $i=1,2$. We take the simplification one step further by showing that message splitting at the broadcast transmitter is also unnecessary under the two conditions. This is accomplished by proving the equivalence of $\mathcal{R}_i$ and $\mathcal{R}_{(i)}$ for $i=1,2$, by inspecting their dominant extreme points, where $\mathcal{R}_{(i)}$ denote the rate regions when the broadcast transmitter only uses superposition. For the converse, we establish the capacity regions of DM-BICs under the two partial-order broadcast conditions respectively for the strong and very strong interference conditions. It is observed that the strong but not very strong interference condition does not exist when the non-interfered broadcast receiver is interference-cognizant less noisy than the interfered broadcast receiver.

We next investigate a GBIC in detail. For ease of exposition, channel parameters are divided into three regimes. Two of the three regimes are closely related to the two partial-order broadcast conditions, where an ordering of the decodability of the two broadcast receivers exists. Consequently achievable rate regions in these regimes are obtained by specializing $\mathcal{R}_{(i)}$, $i=1,2$, with Gaussian input distributions. For the converse, we use the standard entropy power inequality (EPI) \cite{Cover} and Costa's EPI \cite{NEPI} to deal with weak interference and strong interference respectively. This technique is an extension of the one originated in \cite{Kramer} (also see \cite{Anna_Veer} for discussion of bounds in \cite{Kramer}). In addition, the standard EPI is further used to quantify the trade-off between the rates of the two broadcast receivers, resulting in a bound parameterized by $\alpha$ that resembles the power splitting factor in the achievable scheme. While similar bounding techniques have been used in \cite{Shang II}, compared to \cite{Shang II}, which addressed the broadcasting trade-off in an implicit way, the inclusion of a power splitting factor in our outer bounds reveals the similarity between inner and outer bounds, leading to new capacity results that are not available in \cite{Shang II}, such as Corollary \ref{coro:halfbit} and Theorem \ref{theo:halfbitsumcap}. For the remaining regime, an achievable rate region is given by taking the convex hull of two rate regions, obtained by direct specializations of $\mathcal{R}$. Outer bound similar to the above two regimes is also obtained. We note that these results are not contained in \cite{Shang II} either.

This paper is organized as follows. The channel model is introduced in Section II, followed by the derivation of an achievable rate region for a general DM-BIC in Section III. For DM-BICs with two partial-order broadcast conditions, the equivalence of rate regions is presented in Section IV and capacity results are provided in Section V. Discussions for a GBIC are provided in Section VI and this paper is concluded in Section VII.

\emph{Notation}: We define functions $[x]^+=\max\{x,0\}$, $\mathcal{C}(x)=\frac{1}{2}\log(1+x)$, $\bar{x}=1-x$. We use $\phi$ to denote a constant and $A_{\epsilon}^{(n)}(X,Y)$ to denote the joint typical set of random variables $X^n$ and $Y^n$. We define an indicator function $\mathbbold{1}_{a,b}$: $\mathbbold{1}_{a,b}=1$ if $a=b$ and $\mathbbold{1}_{a,b}=0$ if $a\neq b$. The notation convention follows \cite{Cover}. The logarithm is with base 2.

%%%%%%%%%%%%%%%%%%%%%%%%%%%%%%%%%%%%%%%%%%%%%____Channel Model_____%%%%%%%%%%%%%%%%%%%%%%%%%%%%%%%%%%%%%%%%%%%%%%%
\section{Channel Model}
\begin{figure}[htb]
    \centering
    \includegraphics[width=95mm]{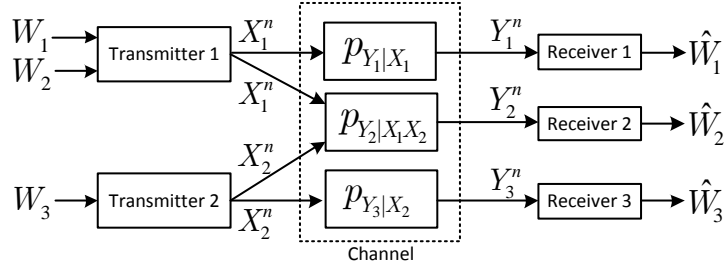}
    \caption{Discrete memoryless broadcast interference channel}
    \label{ChnlMo}
\end{figure}

A discrete memoryless broadcast interference channel is denoted by $(\mathcal{X}_1\times \mathcal{X}_2 ,p(y_1,y_2,y_3|x_1,x_2), \mathcal{Y}_1\times\mathcal{Y}_2\times\mathcal{Y}_3)$, where $\mathcal{X}_i$, $i=1,2$, are the input alphabets, $\mathcal{Y}_j$, $j=1,2,3$, are the output alphabets and $p(y_1,y_2,y_3|x_1,x_2)$ is the channel transition probability. In this paper, we concentrate on a specific interference pattern, where $p(y_1,y_2,y_3|x_1,x_2)= p(y_1|x_1)p(y_2|x_1,x_2)p(y_3|x_2)$. As shown in Fig \ref{ChnlMo}, while transmitter 1 wishes to broadcast to receivers 1, 2 , the second receiver is interfered by transmitter 2 who wishes to communicate with receiver 3.

\begin{definition}
A $(M_1,M_2,M_3,n)$ code consists of message sets $\mathcal{W}_j=\{1,...,M_j\}$; two encoding functions $X_1: (\mathcal{W}_1\times\mathcal{W}_2) \rightarrow \mathcal{X}_1^n$, $X_2:\mathcal{W}_3\rightarrow \mathcal{X}_2^n$ and three decoding functions $g_j:\mathcal{Y}_j^n\rightarrow \mathcal{W}_j$, $j=1,2.3$.
\end{definition}

The messages $W_j$ are uniformly distributed on $\mathcal{W}_j$, where the average error probability for the $(M_1,M_2,M_3,n)$ code is
\begin{equation*}
    P_e=\textrm{Pr}(g_1(Y_1^n)\neq W_1 \textrm{ or }g_2(Y_2^n)\neq W_2 \textrm{ or }g_3(Y_3^n)\neq W_3).
\end{equation*}

\begin{definition}
Rates of a $(M_1,M_2,M_3,n)$ code are defined as $R_j=\frac{\log(M_j)}{n}$ for $j=1,2,3$.
\end{definition}

Rates $(R_1,R_2,R_3)$ are said to be \textit{achievable} if there exists a sequence of $(M_1,M_2,M_3,n)$ codes with $P_e\rightarrow 0$ as $n\rightarrow \infty$. An achievable rate region is the set of all achievable rates for a given coding scheme. The capacity region is the closure of the union of all achievable rate regions.

\begin{figure}[htb]
    \centering
    \includegraphics[width=65mm]{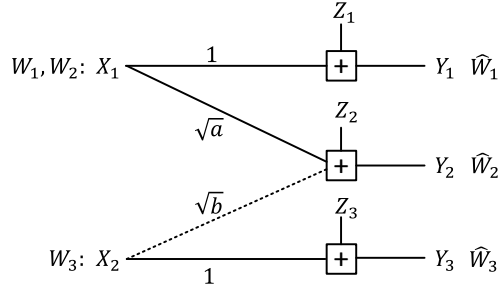}
    \caption{Gaussian broadcast interference channel}
    \label{GBICChnl}
\end{figure}

The Gaussian broadcast interference channel studied in this paper, depicted in Fig \ref{GBICChnl}, is given by
\begin{subequations}
\label{channel}
\begin{align}
    Y_1&=X_1+Z_1 \\
    Y_2&=\sqrt{a}X_1 + \sqrt{b}X_2 + Z_2  \\
    Y_3&=X_2+Z_3,
\end{align}
\end{subequations}
where $Z_i\sim\mathcal{N}(0,1)$ for $i\in\{1,2,3\}$ is the i.i.d. Gaussian noise process and the inputs are subject to power constraints: %$\frac{1}{n}\sum_{i=1}^n x_i^2\leq P_i$ for $i\in\{1,2\}$.
$E(X_i^2)\leq P_i$, $i\in\{1,2\}$. Note that a GBIC with arbitrary channel parameters can always be transformed to the above form.

%%%%%%%%%%%%%%%%%%%%%%%%%%%%%%%%%%%%%%%%%%%%%____Achievable Rate Regions_____%%%%%%%%%%%%%%%%%%%%%%%%%%%%%%%%%%%%%%%
\section{Achievable Rate Region for a General DM-BIC}
In this section, we derive an achievable rate region for the general DM-BIC, where the broadcast transmitter employs message splitting, superposition and binning and the interfering transmitter employs message splitting and superposition.

\begin{theorem}
\label{BinningRegion}
Let $\mathcal{P}$ be the set of joint input probability distributions $P$ that factor as
\begin{align}
  P&=P_{QU_1V_1V_2U_2X_2}(q,u_1,v_1,v_2,u_2,x_2) \notag \\
&=P_{Q}(q)P_{U_1|Q}(u_1|q)P_{V_1V_2|U_1Q}(v_1,v_2|u_1,q)P_{U_2|Q}(u_2|q)P_{X_2|U_2Q}(x_2|u_2,q). \label{pdffactor}
\end{align}
For a given $P\in\mathcal{P}$ and some function $X_1=f(U_1,V_1,V_2)$, let $\mathcal{R}_P$ be the set of non-negative $(R_1,R_2,R_3)$ satisfying
\begin{align}
    R_1&\leq I(V_1;Y_1|Q)\label{ineq1}\\
    R_2&\leq I(V_2;Y_2|U_2,Q)\label{ineq2}\\
    R_3&\leq I(X_2;Y_3|Q)\label{ineq3}\\
    R_1+R_2&\leq I(V_1;Y_1|U_1,Q) + I(V_2;Y_2|U_2,Q)- I(V_1;V_2|U_1,Q)\label{ineq4}\\
    R_1+R_2&\leq I(V_1;Y_1|Q) + I(V_2;Y_2|U_1,U_2,Q)- I(V_1;V_2|U_1,Q)\label{ineq5}\\
    R_2+R_3&\leq I(V_2,U_2;Y_2|Q)+I(X_2;Y_3|U_2,Q)\label{R2plusR3}\\
    R_1+R_2+R_3&\leq I(V_1;Y_1|U_1,Q)+I(V_2,U_2;Y_2|Q)+ I(X_2;Y_3|U_2,Q)-I(V_1;V_2|U_1,Q)\label{sum1}\\
    R_1+R_2+R_3&\leq I(V_1;Y_1|Q)+I(V_2,U_2;Y_2|U_1,Q)+ I(X_2;Y_3|U_2,Q)-I(V_1;V_2|U_1,Q)\label{sum2}.
\end{align}
Then the region $\mathcal{R}=\bigcup_{P\in\mathcal{P}} \mathcal{R}_P$ is an achievable rate region for a DM-BIC.
\end{theorem}

\begin{remark}
The main idea for the achievability proof of $\mathcal{R}$ can be summarized as follows. The messages for receivers 1 and 2 are split into common and private parts respectively. Common messages are carried by the cloud center $U_1$, which is decoded at both $Y_1$ and $Y_2$. The private message carriers $V_1$ and $V_2$, which are only decoded at their respective intended receivers, are superimposed upon $U_1$, where binning is used to generate arbitrary dependence between $V_1$ and $V_2$. Similarly at the interfering transmitter, message splitting and superposition are employed. As a result, the common message is encoded into the cloud center $U_2$ and the satellite signal $X_2$ bears both common and private messages. Each receiver uses a joint-typicality decoder.
\end{remark}

\begin{remark}
With $X_2=U_2=U_1=\phi$ and $R_3=0$, $\mathcal{R}$ reduces to the Marton's region with private message sets for a DM-BC \cite{El Gamal}. With $U_1=V_1=\phi$, $X_1=V_2$ and $R_1=0$, $\mathcal{R}$ reduces to the compact Han-Kobayashi region \cite{CMG} for a one-sided interference channel.
\end{remark}

\subsection{Proof of Theorem \ref{BinningRegion}}
We will prove Theorem \ref{BinningRegion} in several steps. First, we obtain an achievable rate region $\hat{\mathcal{R}}$ in Lemma \ref{R3hat} using message splitting, superposition and binning. Then we simplify region $\hat{\mathcal{R}}$ to region $\tilde{\mathcal{R}}$ in Proposition \ref{prop:simplify1} by removing certain restrictions on the input distribution. Finally in Proposition \ref{equivalence}, we simplify region $\tilde{\mathcal{R}}$ to region $\mathcal{R}$, where redundant inequalities are removed.

\begin{lemma}
\label{R3hat}
Let $\hat{\mathcal{P}}$ be the set of joint input probability distributions $P$ that factor as in (\ref{pdffactor}) with the additional constraint that
\begin{align}
    I(V_1;Y_1|U_1,Q)+I(V_2;Y_2|U_1,U_2,Q)-I(V_1;V_2|U_1,Q)\geq 0. \label{cnst1}
\end{align}
For a given $P\in\hat{\mathcal{P}}$ and some function $X_1=f(U_1,V_1,V_2)$, let $\hat{\mathcal{R}}_P$ be the set of non-negative $(R_1,R_2,R_3)$ satisfying inequalities (\ref{ineq1})-(\ref{sum2}), plus the following:
\begin{align}
    R_3&\leq I(V_2,U_2;Y_2|U_1,Q) + I(X_2;Y_3|U_2,Q)\label{rdt1}\\
    R_3&\leq I(V_1;Y_1|U_1,Q) + I(V_2,U_2;Y_2|U_1,Q) + I(X_2;Y_3|U_2,Q)- I(V_1;V_2|U_1,Q)\label{rdt2}.
\end{align}
Then the region $\hat{\mathcal{R}}=\bigcup_{P\in\hat{\mathcal{P}}} \hat{\mathcal{R}}_P$ is an achievable rate region for a DM-BIC.
\end{lemma}

\begin{IEEEproof}
See Appendix \ref{ProofLemmaR3hat}.
\end{IEEEproof}

Let us denote $\tilde{\mathcal{R}}=\bigcup_{P\in\mathcal{P}}\hat{\mathcal{R}}_P$. Since $\hat{\mathcal{P}}\subseteq\mathcal{P}$ and $\hat{\mathcal{R}}_P\subseteq\mathcal{R}_P$, we have
\begin{equation*}
\bigcup_{P\in\hat{\mathcal{P}}}\hat{\mathcal{R}}_P\subseteq \bigcup_{P\in\mathcal{P}}\hat{\mathcal{R}}_P \subseteq \bigcup_{P\in\mathcal{P}}\mathcal{R}_P,
\end{equation*}
i.e. $\hat{\mathcal{R}}\subseteq\tilde{\mathcal{R}}\subseteq\mathcal{R}$. In the following, we prove that these three regions are essentially equivalent and hence $\mathcal{R}$ is also achievable.

%Comparing $\mathcal{R}$ with $\hat{\mathcal{R}}$, we notice that $\mathcal{P}\supseteq\hat{\mathcal{P}}$ and $\mathcal{R}_P\supseteq\hat{\mathcal{R}}_P$. If we further define $\tilde{\mathcal{R}}=\bigcup_{P\in\mathcal{P}}\hat{\mathcal{R}}_P$, then we have $\mathcal{R}\supseteq\tilde{\mathcal{R}}\supseteq\hat{\mathcal{R}}$. In the following, we prove that these three regions are essentially equivalent and hence $\mathcal{R}$ is also achievable.
%In what follows, we will show that the constraint (\ref{cnst1}) on input distributions is unnecessary. First of all, (\ref{cnst1}) is a direct consequence of the nonnegativity of some intermediate rates, which are eventually eliminated using Fourier-Motzkin procedure. To be more precise, in Appendix \ref{ProofLemmaR3hat} (\ref{cnstinput1}), (\ref{cnstinput2}), and $R_1'+R_2'\geq I(V_1;V_2|U_1)$ entails (\ref{cnst1}) for any given input distribution with fixed $Q$. However, if we consider the union of regions corresponding to all input distributions, the region evaluated for a non-conforming input distribution will be contained within those of conforming input distributions. Hence if our interest is only in the rate region boundary as we do here, the constraint (\ref{cnst1}) is unnecessary.

\begin{proposition}
\label{prop:simplify1}
For $\hat{\mathcal{R}}=\bigcup_{P\in\hat{\mathcal{P}}}\hat{\mathcal{R}}_P$ and $\tilde{\mathcal{R}}=\bigcup_{P\in\mathcal{P}}\hat{\mathcal{R}}_P$, where $\mathcal{P}$ is defined in Theorem \ref{BinningRegion} and $\hat{\mathcal{P}}$, $\hat{\mathcal{R}}_P$ are defined in Lemma \ref{R3hat}, we have $\hat{\mathcal{R}}=\tilde{\mathcal{R}}$.
\end{proposition}

\begin{IEEEproof}
If we define $\bar{\mathcal{P}}=\mathcal{P}\setminus\hat{\mathcal{P}}$, then $\tilde{\mathcal{R}}= \left(\bigcup_{P\in\hat{\mathcal{P}}} \hat{\mathcal{R}}_{P}\right) \bigcup \left( \bigcup_{P\in\bar{\mathcal{P}}} \hat{\mathcal{R}}_P \right)=\hat{\mathcal{R}} \bigcup \left( \bigcup_{P\in\bar{\mathcal{P}}} \hat{\mathcal{R}}_P \right) $. Next we prove that for any $P\in\bar{\mathcal{P}}$, there exists a $P'\in\hat{\mathcal{P}}$ such that $\hat{\mathcal{R}}_P\subseteq\hat{\mathcal{R}}_{P'}$. Therefore $\tilde{\mathcal{R}}=\hat{\mathcal{R}}$.

For a given $P\in\bar{\mathcal{P}}$ that does not satisfy (\ref{cnst1}), from Appendix \ref{ProofLemmaR3hat} it is clear that $\hat{\mathcal{R}}_{P}$ is not achievable using the coding scheme for Lemma \ref{R3hat}. Nevertheless if we evaluate $\hat{\mathcal{R}}_{P}$ for this $P$, the resulting region is then contained within an outer bound consisting the following inequalities% and recall that
%\begin{align*}
%  I(V_1;Y_1|U_1,Q)+I(V_2;Y_2|U_1,U_2,Q)-I(V_1;V_2|U_1,Q)<0.
%\end{align*}
%After some manipulation, we can show that $\hat{\mathcal{R}}_{P}$ is contained within the following region
\begin{align}
  R_3&\leq I(X_2,Y_3|Q) \label{rhat_1}\\
  R_3&\leq I(U_2;Y_2|U_1,Q) + I(X_2;Y_3|U_2,Q) \label{rhat_2}\\
  R_1+R_2&\leq I(U_1;Y_1|Q) \label{rhat_3}\\
  R_1+R_2&\leq I(U_1;Y_2|U_2,Q) \label{rhat_4}\\
  R_1+R_2+R_3&\leq I(U_1,U_2;Y_2|Q) + I(X_2;Y_3|U_2,Q) \label{rhat_5}\\
  R_1,R_2,R_3&\geq 0. \label{rhat_6}
\end{align}
To see this, note that the right-hand side of \eqref{rdt2} can be written as
\begin{align*}
    I(U_2;Y_2|U_1,Q) + I(X_2;Y_3|U_2,Q) + I(V_1;Y_1|U_1,Q) + I(V_2;Y_2|U_1,U_2,Q) - I(V_1;V_2|U_1,Q).
\end{align*}
Recall that $I(V_1;Y_1|U_1,Q)+I(V_2;Y_2|U_1,U_2,Q)-I(V_1;V_2|U_1,Q)<0$. Hence we have \eqref{rhat_2}. Similarly, we can derive \eqref{rhat_3} from \eqref{ineq5}, \eqref{rhat_4} from \eqref{ineq4}, \eqref{rhat_5} from \eqref{sum1}. For the given $P$ above, let us consider another probability distribution function $P'$ which differs from $P$ only by
\begin{equation*}
    P_{V_1V_2|U_1Q}(v_1,v_2|u_1,q)=\begin{cases}
        1,\quad v_1=v_2=u_1,\\
        0,\quad\text{otherwise}
    \end{cases}.
\end{equation*}
%$P_{V_1|U_1Q}(v_1|u_1,q)=\mathbbold{1}_{v_1,u_1}$ and $P_{V_2|U_1,Q}(v_2|u_1,q)=\mathbbold{1}_{v_2,u_1}$ for any $q$, where
%\begin{align}
%\label{indicator}
%  \mathbbold{1}_{a,b}=\begin{cases}
%                     1,\quad a=b \\
%		     0,\quad a\neq b
%                    \end{cases}.
%\end{align}
It can be checked that $P'\in\hat{\mathcal{P}}$ and the above region defined by \eqref{rhat_1}--\eqref{rhat_6} is exactly $\hat{\mathcal{R}}_{P'}$. Hence $\mathcal{R}_{P} \subseteq  \hat{\mathcal{R}}_{P'}$.
\end{IEEEproof}

Comparing $\mathcal{R}=\bigcup_{P\in\mathcal{P}} \mathcal{R}_P$ with $\tilde{\mathcal{R}}=\bigcup_{P\in\mathcal{P}} \hat{\mathcal{R}}_P$, we see that $\hat{\mathcal{R}}_P$ contains two more inequalities, (\ref{rdt1}) and (\ref{rdt2}), than $\mathcal{R}_P$, resulting in $\tilde{\mathcal{R}}\subseteq\mathcal{R}$. We next show that these two inequalities are redundant and establish the equivalence of $\mathcal{R}$ and $\tilde{\mathcal{R}}$ and consequently $\mathcal{R}=\hat{\mathcal{R}}$.

\begin{proposition}
\label{equivalence}
$\mathcal{R}=\tilde{\mathcal{R}}$, where $\mathcal{R}$ is defined in Theorem \ref{BinningRegion} and $\tilde{\mathcal{R}}$ is defined in Lemma \ref{R3hat}.
\end{proposition}

\begin{IEEEproof}
In the following, we use a technique that has been previously used in \cite{CMG} to simplify the Han-Kobayashi region. For a given $P\in\mathcal{P}$, let $P_{U_2=\phi}$ denote a probability distribution that differs from $P$ only by $P_{U_2|Q}(u_2|q)=\mathbbold{1}_{u_2,\phi}$ for any $q$, where $\phi$ is a constant and $\mathbbold{1}_{u_2,\phi}$ is an indicator function defined in Introduction. To prove (\ref{rdt1}) is redundant, we will show that if $\mathbf{R}=(R_1,R_2,R_3)$ satisfies all inequalities defining $\hat{\mathcal{R}}_P$ except (\ref{rdt1}), then $\mathbf{R}\in\hat{\mathcal{R}}_{P'}$, where $P'=P_{U_2=\phi}$. Hence by the union operation in $\mathcal{R}$ and $\tilde{\mathcal{R}}$, (\ref{rdt1}) is redundant.

If (\ref{rdt1}) is violated, we have
\begin{align}
    R_3> I(V_2,U_2;Y_2|U_1,Q) + I(X_2;Y_3|U_2,Q). \label{violated1}
\end{align}
If $\mathbf{R}$ satisfies all inequalities defining $\hat{\mathcal{R}}$ except (\ref{rdt1}), then $\mathbf{R}$ satisfies the following
\begin{align}
    R_1&\leq I(V_1;Y_1|Q)\notag\\
    R_2&\leq I(U_1;Y_2|Q)\label{equi1}\\
    R_3&\leq I(X_2;Y_3|Q)\notag\\
    R_1+R_2&\leq I(V_1;Y_1|U_1,Q) + I(U_1;Y_2|Q) - I(V_1;V_2|U_1,Q)\label{equi2}\\
    R_1+R_2&\leq I(V_1;Y_1|Q) - I(V_1;V_2|U_1,Q)\label{equi3}\\
    R_1,R_2,R_3&\geq 0,\notag
\end{align}
where (\ref{equi1}) is obtained from (\ref{R2plusR3}) and (\ref{violated1}), (\ref{equi2}) from (\ref{sum1}) and (\ref{violated1}), (\ref{equi3}) from (\ref{sum2}) and (\ref{violated1}). Note that (\ref{rdt2}) and (\ref{violated1}) ensure that the right-hand sides of (\ref{equi2}) and (\ref{equi3}) are positive. It is easy to show that the above region is contained within $\hat{\mathcal{R}}_{P'}$. Therefore (\ref{rdt1}) is redundant. In the following we assume (\ref{rdt1}) has already been removed.

Similarly if (\ref{rdt2}) is violated, we have
\begin{align}
    R_3&> I(V_1;Y_1|U_1,Q) + I(V_2,U_2;Y_2|U_1,Q) + I(X_2;Y_3|U_2,Q)- I(V_1;V_2|U_1,Q). \label{violated2}
\end{align}
Hence if $\mathbf{R}$ satisfies all inequalities defining $\hat{\mathcal{R}}_P$ except (\ref{rdt2}), then $\mathbf{R}$ satisfies the following
\begin{align}
  R_1+R_2&\leq I(U_1;Y_2|Q) \label{equi22}\\
  R_1+R_2&\leq I(U_1;Y_1|Q) \label{equi11}\\
  R_3&\leq I(X_2;Y_3|Q)\notag\\
  R_1,R_2,R_3&\geq 0,\notag
\end{align}
where (\ref{equi22}) is from (\ref{sum1}) and (\ref{violated2}) and (\ref{equi11}) is from (\ref{sum2}) and (\ref{violated2}). This region is clearly contained within $\hat{\mathcal{R}}_{P'}$. Hence (\ref{rdt2}) is also redundant.
\end{IEEEproof}

%%%%%%%%%%%%%%%%%%%%%%%%%%%%%%%%%%%%%%%%%%%%%%%%____Equivalence_____%%%%%%%%%%%%%%%%%%%%%%%%%%%%%%%%%%%%%%%%%%%%%%%
\section{Achievable Rate Regions of DM-BICs under Partial-Order Broadcast Conditions}
In this section, we concentrate on DM-BICs under two partial-order broadcast conditions: interference-oblivious less noisy and interference-cognizant less noisy, which are defined as follows.

\begin{definition}
In a DM-BIC, receiver 1 is said to be {\em interference-cognizant less noisy} than receiver 2, denoted by $Y_1\succ_c Y_2$, if $I(U_1;Y_1)\geq I(U_1;Y_2|X_2)$ for all $p(u_1,x_1)p(x_2)$ such that $U_1\rightarrow (X_1,X_2)\rightarrow (Y_1,Y_2)$ form a Markov chain.
\end{definition}

\begin{definition}
In a DM-BIC, receiver 2 is said to be {\em interference-oblivious less noisy} than receiver 1, denoted by $Y_1\prec_o Y_2$, if $I(U_1;Y_1)\leq I(U_1;Y_2)$ for all $p(u_1,x_1)p(x_2)$ such that $U_1\rightarrow (X_1,X_2)\rightarrow (Y_1,Y_2)$ form a Markov chain.
\end{definition}

\begin{definition}
In a DM-BIC, receiver 2 is said to be {\em physically degraded} with respect to receiver 1 if $p(y_2|x_1,y_1,x_2)=p(y_2|y_1,x_2)$, i.e. $X_1\rightarrow (Y_1,X_2)\rightarrow Y_2$ form a Markov chain for all $p(x_1)p(x_2)$.
\end{definition}

\begin{definition}
In a DM-BIC, receiver 2 is said to be {\em stochastically degraded} with respect to receiver 1 if there exists a random variable $\tilde{Y}_1$ such that $p(\tilde{y}_1|x_1)=p(y_1|x_1)$ and $p(y_2|x_1,\tilde{y}_1,x_2)=p(y_2|\tilde{y}_1,x_2)$, i.e. $X_1\rightarrow (\tilde{Y}_1,X_2)\rightarrow Y_2$ form a Markov chain for all $p(x_1)p(x_2)$. Similarly receiver 1 is stochastically degraded with respect to receiver 2 if there exists a random variable $\tilde{Y}_2$ such that $p(\tilde{y}_2|x_1)=p(y_2|x_1)$ and $p(y_1|x_1,\tilde{y}_2)=p(y_1|\tilde{y}_2)$, i.e. $X_1\rightarrow \tilde{Y}_2\rightarrow Y_1$ form a Markov chain for all $p(x_1)$.
\end{definition}

\begin{remark}
We can interpret $Y_1\succ_c Y_2$ as follows: even without the presence of interference, i.e. $X_2$ is provided to receiver 2, receiver 1 is still less noisy. Reversely, $Y_1\prec_o Y_2$ says that even if no particular action is taken by receiver 2 to deal with interference, receiver 2 is still {\em less noisy} than receiver 1 \cite{Korner}. Note that both physical and stochastic degradedness
%, which are equivalent in this case \cite[Theorem 15.6.1]{Cover})
implies the partial-order conditions and hence are stricter. In particular, $Y_2$ being degraded (either physically or stochastically) with respect to $Y_1$ implies $Y_1\succ_c Y_2$, but not vice versa. Also note that $Y_1$ in general cannot be physically degraded with respect to $Y_2$, since receiver 1 is not subject to interference.
\end{remark}

The first class of schemes we consider here is a specialization of $\mathcal{R}$ in Theorem \ref{BinningRegion} for the two partial-order broadcast conditions, where binning is removed. We call the resulting regions as $\mathcal{R}_i$, $i=1,2$. The rationale of such specialization is that for a regular broadcast channel, binning is unnecessary when there exists certain ordering among receivers, such as having degraded, less noisy and more capable condition \cite{El Gamal}. We expect the same for a DM-BIC. However due to the complication of interference, we are only able to prove the optimality of $\mathcal{R}_i$, $i=1,2$, for certain channel conditions, which will be shown in Section V.

\begin{corollary}
\label{CoroR1}
Let $\mathcal{P}'$ be the set of joint input probability distributions $P$ that factor as
\begin{align}
  P=P_{U_1X_1U_2X_2}(u_1,x_1,u_2,x_2)=P_{U_1}(u_1)P_{X_1|U_1}(x_1|u_1)P_{U_2}(u_2)P_{X_2|U_2}(x_2|u_2)\label{newprop}
\end{align}
For a given $P\in\mathcal{P}'$, let $\mathcal{R}_{1,P}$ be the set of non-negative $(R_1,R_2,R_3)$ satisfying
\begin{align}
    R_1&\leq I(U_1;Y_1)\label{ts1}\\
    R_3&\leq I(X_2;Y_3)\label{ts2}\\
    R_1+R_2&\leq I(U_1;Y_1)+I(X_1;Y_2|U_1,U_2) \label{largestR22}\\
    R_1+R_2+R_3&\leq I(U_1;Y_1)+I(X_1,U_2;Y_2|U_1)+ I(X_2;Y_3|U_2). \label{largestR33}
\end{align}
Then $\mathcal{R}_1=\bigcup_{P\in\mathcal{P}'}\mathcal{R}_{1,P}$ is an achievable rate region for a DM-BIC with $Y_1\prec_o Y_2$.
\end{corollary}

\begin{IEEEproof}
Let $Q=\phi$. Specializing $\mathcal{R}_P$ in Theorem \ref{BinningRegion} with $V_2=X_1$, $V_1=U_1$ and removing redundant inequalities due to $Y_1\prec_o Y_2$, we obtain $\mathcal{R}_{1,P}$. By taking the union of $\mathcal{R}_{1,P}$ for all $P\in\mathcal{P}'$, we obtain $\mathcal{R}_1$.
\end{IEEEproof}

\begin{corollary}
\label{CoroR2}
For a given $P\in\mathcal{P}'$, where $P$ factors as (\ref{newprop}), let $\mathcal{R}_{2,P}$ be the set of non-negative $(R_1,R_2,R_3)$ satisfying
\begin{align}
    R_2&\leq I(U_1;Y_2|U_2)\label{largestR2}\\
    R_3&\leq I(X_2;Y_3)\notag\\
    R_1+R_2&\leq I(X_1;Y_1|U_1)+I(U_1;Y_2|U_2)\label{largestR1}\\
    R_2+R_3&\leq I(U_1,U_2;Y_2)+I(X_2;Y_3|U_2)\label{largestR3}\\
    R_1+R_2+R_3&\leq I(X_1;Y_1|U_1)+I(U_1,U_2;Y_2)+I(X_2;Y_3|U_2).\notag
\end{align}
Then $\mathcal{R}_{2}=\bigcup_{P\in\mathcal{P}'}\mathcal{R}_{2,P}$ is an achievable rate region for a DM-BIC with $Y_1\succ_c Y_2$.
\end{corollary}

\begin{IEEEproof}
A direct specialization of $\mathcal{R}$ in Theorem \ref{BinningRegion} will result in some extra inequalities that are difficult to remove. Hence instead, the specialization will be done for an equivalent region of $\mathcal{R}$. The proof is provided in Appendix \ref{ProofCoroR2}.
\end{IEEEproof}

Note that to derive $\mathcal{R}_i$, we fix the time-sharing random variable $Q$. In principle, we could have kept $Q$ intact when specializing $\mathcal{R}$, but the following proposition asserts that there is no benefit doing so. Since time-sharing always results in a region no smaller than the convex hull operation \cite{El Gamal}, it follows that taking convex hull is also unnecessary.

\begin{proposition}
\label{convexification}
$\mathcal{R}_i$, $i=1,2$, is not enlarged by using time-sharing.
\end{proposition}

\begin{IEEEproof}
See Appendix \ref{ProofPropositionConv}.
\end{IEEEproof}

Next we present two achievable rate regions, $\mathcal{R}_{(i)}$, $i=1,2$, where the broadcast transmitter uses only superposition coding with the cloud center carrying only user $i$'s message. Since the proofs are standard, they are omitted for conciseness.

\begin{proposition}
\label{Rprime}
For a given $P\in\mathcal{P}'$, where $P$ factors as (\ref{newprop}), let $\mathcal{R}_{(1),P}$ be the set of non-negative $(R_1,R_2,R_3)$ satisfying
\begin{align}
    R_1&\leq I(U_1;Y_1)\\ %\label{Rprime1}\\
    R_2&\leq I(X_1;Y_2|U_1,U_2)\\%\label{Rprime2}\\
    R_3&\leq I(X_2;Y_3)\\%\notag\\
    R_2+R_3&\leq I(X_1,U_2;Y_2|U_1)+I(X_2;Y_3|U_2),%\label{Rprime4}\\
\end{align}
Then $\mathcal{R}_{(1)}=\bigcup_{P\in\mathcal{P}'}\mathcal{R}_{(1),P}$ is an achievable rate region for a DM-BIC with $Y_1\prec_o Y_2$.
\end{proposition}

\begin{proposition}
\label{Rpprime}
For a given $P\in\mathcal{P}'$, where $P$ factors as (\ref{newprop}), let $\mathcal{R}_{(2),P}$ be the set of non-negative $(R_1,R_2,R_3)$ satisfying
\begin{align}
    R_1&\leq I(X_1;Y_1|U_1)\\%\label{Rpprime1}\\
    R_2&\leq I(U_1;Y_2|U_2)\\%\label{Rpprime2} \\
    R_3&\leq I(X_2;Y_3)\\%\notag\\
    R_2+R_3&\leq I(U_1,U_2;Y_2)+I(X_2;Y_3|U_2),%\notag\\
\end{align}
Then $\mathcal{R}_{(2)}=\bigcup_{P\in\mathcal{P}'}\mathcal{R}_{(2),P}$ is an achievable rate region for a DM-BIC with $Y_1\succ_c Y_2$.
\end{proposition}

In the derivation of $\mathcal{R}$, we used message splitting, superposition and binning at the broadcast transmitter. Regions $\mathcal{R}_i$, $i=1,2,$ are derived from $\mathcal{R}$ when binning is stripped off but message splitting and superposition are kept intact. While both $\mathcal{R}_i$ and $\mathcal{R}_{(i)}$ rely on superposition coding at the broadcast transmitter, there is a subtle difference. Despite the fact that both schemes' cloud centers carry receiver $i$'s message, the one for $\mathcal{R}_i$ could in addition carry receiver $j$'s ($j=1,2,\ j\neq i$) common message, which might be helpful to reduce self-interference due to the fact that part of the broadcast signal intended for receiver $j$ is essentially interference from receiver $i$'s perspective. Also it is apparent that the rate regions based only on superposition at the broadcast transmitter are no larger than the ones based on both superposition and message splitting, which can be also verified by checking that the inequalities defining $\mathcal{R}_{(i)}$ induce those in $\mathcal{R}_i$, but not vice versa. Hence it seems that $\mathcal{R}_i$ is strictly larger than $\mathcal{R}_{(i)}$. However, if we consider the no interference case with $U_2=X_2=\phi$, $R_3=0$, i.e. a regular DM-BC, $\mathcal{R}_i$ cannot be strictly larger than $\mathcal{R}_{(i)}$ since the latter is the capacity region of a less noisy DM-BC \cite{El Gamal}. The pitfall of the previous argument is that it only considers a specific input distribution. It is true that for some $P\in\mathcal{P}'$, $\mathcal{R}_{i,P}$ defined in Corollary \ref{CoroR1} and \ref{CoroR2} are strictly larger than $\mathcal{R}_{(i),P}$ defined in Proposition \ref{Rprime} and \ref{Rpprime} respectively, however once we consider all input distributions, the regions $\mathcal{R}_i$ and $\mathcal{R}_{(i)}$ are indeed equivalent as shown in the following theorem.

\begin{theorem}
\label{theoremequi}
$\mathcal{R}_i=\mathcal{R}_{(i)}$ for $i=1,2$, where $\mathcal{R}_i$ are given in Corollary \ref{CoroR1} and \ref{CoroR2} respectively and $\mathcal{R}_{(i)}$ are given in Proposition \ref{Rprime} and \ref{Rpprime} respectively.
\end{theorem}

%____________________________________________________________________________________________________________________
\subsection{Proof of Theorem \ref{theoremequi}}
Before proving Theorem \ref{theoremequi}, we need the following definitions and lemmas.

\begin{definition}
Let $\mathbb{R}_c^n$ be a convex subset of $\mathbb{R}^n$. A point $X\in\mathbb{R}_c^n$ is an \textit{extreme point (ExP)} iff whenever $X=tY+(1-t)Z$, $t\in(0,1)$ and $Y\neq Z$, this implies either $Y\not \in \mathbb{R}_c^n$ or $Z\not \in \mathbb{R}_c^n$.
\end{definition}

\begin{definition}
Let $\mathbb{R}_c^n$ be a convex subset of $\mathbb{R}^n$. A point $X\in\mathbb{R}_c^n$ is a \textit{dominant extreme point (DExP)} iff $X$ is an ExP and there does not exist another ExP $Y\in\mathbb{R}_c^n$, $Y\neq X$, such that $X\leq Y$ element-wise.
\end{definition}

\begin{remark}
In the literature, the term ``dominant extreme points'' are sometimes referred as corner points. The intention of choosing the former terminology is to emphasize the connection to convex set.
\end{remark}

Let $\mathcal{R}^n$ be a $n$-dimensional convex rate region and $\Omega(\mathcal{R}^n)$ be the set of all DExPs of $\mathcal{R}^n$. Furthermore let $co(\cdot)$ denote the convex hull operation. In particular
\begin{align*}
  co(\Omega)=\left\{\sum_{i=1}^{m}\alpha_i\mathbf{R}_i :\mathbf{R}_i \in\Omega, \alpha_i\in[0,1], \sum_{i=1}^m\alpha_i=1, m=1,2,... \right\}.
\end{align*}

\begin{lemma}
\label{DominateEx}
$\mathbf{R}\in\mathcal{R}^n$ iff there exists some $\mathbf{R}'\in co(\Omega)$ such that $\mathbf{R}\leq \mathbf{R}'$ element-wise.
\end{lemma}

\begin{IEEEproof}
For the ``if'' part, since DExPs are achievable, so are their convex combinations, specifically $\mathbf{R}'$ is achievable. If $\mathbf{R}\leq \mathbf{R}'$, then $\mathbf{R}$ is also achievable. For the ``only if'' part, since $\mathcal{R}^n$ is a convex region, any point in $\mathcal{R}^n$ can be expressed as a convex combination of its ExPs, i.e. there exists some $\mathbf{R}_i \in\Psi$, $\alpha_i\in[0,1]$ and an integer $m$ such that $\mathbf{R}=\sum_{i=1}^m \alpha_i\mathbf{R}_i$, where $\Psi$ denotes the set of all ExPs of $\mathcal{R}^n$. Now replacing any non-dominant $\mathbf{R}_i$ that constitutes $\mathbf{R}$ by its corresponding DExP and keeping convex coefficients $\alpha_i$ intact, we obtain $\mathbf{R}'\in co(\Omega)$, where $\mathbf{R}\leq \mathbf{R}'$.
\end{IEEEproof}

Lemma \ref{DominateEx} suggests that a rate region is completely specified by its DExPs. Hence the key to prove $\mathcal{R}_i=\mathcal{R}_{(i)}$ is to find their DExPs. This is accomplished by first finding the DExPs of the constituent regions $\mathcal{R}_{i,P}$ and $\mathcal{R}_{(i),P}$ as shown in the following lemmas.

\begin{lemma}
\label{DEXPR2minus}
For a DM-BIC with $Y_1\succ_c Y_2$ and a given $P\in\mathcal{P}'$, where $P$ factors as (\ref{newprop}), the set of DExPs of $\mathcal{R}_{2,P}$ in Corollary \ref{CoroR2} is given by $\Omega(\mathcal{R}_{2,P})=\{A,B,C,D\}$, where
\begin{align*}
    A &= (\ I(X_1;Y_1|U_1),\ I(U_1;Y_2|U_2),\ \min\{ I(X_2;Y_3),\ I(U_2;Y_2)+ I(X_2;Y_3|U_2) \} \ ),\\
    B &= (\ I(X_1;Y_1|U_1),\ \min\{I(U_1;Y_2|U_2),\ [I(U_1,U_2;Y_2)-I(U_2;Y_3)]^+ \} ,\ I(X_2;Y_3|U_2)+ \\
    &\qquad  \min\{ I(U_2;Y_3) ,\ I(U_1,U_2;Y_2) \} \ ),\\
    C &= (\ I(X_1;Y_1|U_1)+ \min\{I(U_1;Y_2|U_2),\ [I(U_1,U_2;Y_2)-I(U_2;Y_3)]^+ \}, \ 0 ,\ I(X_2;Y_3|U_2)+ \\
    &\qquad  \min\{ I(U_2;Y_3) ,\ I(U_1,U_2;Y_2) \} \ ),\\
    D &= (\ I(X_1;Y_1|U_1)+I(U_1;Y_2|U_2),\ 0 ,\ \min\{ I(X_2;Y_3) ,\  I(U_2;Y_2)+ I(X_2;Y_3|U_2) \} \ ).
\end{align*}
\end{lemma}

\begin{IEEEproof}
See Appendix \ref{proofDEXPR2minus}.
\end{IEEEproof}

\begin{lemma}
\label{DEXPRpprimeminus}
For a DM-BIC with $Y_1\succ_c Y_2$ and a given $P\in\mathcal{P}'$, where $P$ factors as (\ref{newprop}), the set of DExPs of $\mathcal{R}_{(2),P}$ in Proposition \ref{Rpprime} is given by $\Omega(\mathcal{R}_{(2),P})=\{A,B\}$ where $A$ and $B$ are given in Lemma \ref{DEXPR2minus}.
\end{lemma}

\begin{IEEEproof}
See Appendix \ref{proofDEXPRpprimeminus}.
\end{IEEEproof}

\begin{lemma}
\label{DEXPR1minus}
For a DM-BIC with $Y_1\prec_o Y_2$ and a given $P\in\mathcal{P}'$, where $P$ factors as (\ref{newprop}), the set of DExPs of $\mathcal{R}_{1,P}$ in Corollary \ref{CoroR1} is given by
\begin{align*}
  \Omega(\mathcal{R}_{1,P})=\begin{cases}
			      E,F,G,H,I,\quad \text{if } P \text{ satisfies } I(U_2;Y_3)\leq I(U_1;Y_1)+I(X_1,U_2;Y_2|U_1)\\
			      E,F,G,J,\quad \text{if } P \text{ satisfies } I(U_2;Y_3)> I(U_1;Y_1)+I(X_1,U_2;Y_2|U_1)
                            \end{cases},
\end{align*}
where
\begin{align*}
    E&=(\ I(U_1;Y_1),\ I(X_1;Y_2|U_1,U_2),\ \min\{ I(X_2;Y_3),\ I(U_2;Y_2|U_1)+ I(X_2;Y_3|U_2) \} \ ),\\
    F&=(\ I(U_1;Y_1),\ \min\{ I(X_1;Y_2|U_1,U_2),\ [I(X_1,U_2;Y_2|U_1)- I(U_2;Y_3)]^+\} ,\ I(X_2;Y_3|U_2) + \\
    &\qquad  \min\{ I(U_2;Y_3) ,\  I(X_1,U_2;Y_2|U_1) \} \ ),\\
    G&=(\ 0,\ I(U_1;Y_1)+I(X_1;Y_2|U_1,U_2),\ \min\{ I(X_2;Y_3) ,\ I(U_2;Y_2|U_1) +I(X_2;Y_3|U_2)\} \ ),\\
    H &=(\ 0,\ I(U_1;Y_1)+ I(X_1;Y_2|U_1,U_2)+\min\{0,\ I(U_2;Y_2|U_1) -I(U_2;Y_3)\},\  I(X_2;Y_3)\ ),\\
    I &=(\ I(U_1;Y_1) + \min\{ 0 ,\ I(X_1,U_2;Y_2|U_1) - I(U_2;Y_3) \},\ [I(X_1;Y_2|U_1,U_2)+\\
      &\qquad \min\{ 0 ,\ I(U_2;Y_2|U_1)-I(U_2;Y_3)\}]^+,\ I(X_2;Y_3) \} \ ),\\
    J&=(\ 0,\ 0,\ I(U_1;Y_1)+I(X_1,U_2;Y_2|U_1)+I(X_2;Y_3|U_2)\ ).
\end{align*}
\end{lemma}

\begin{IEEEproof}
See Appendix \ref{proofDEXPR1minus}.
\end{IEEEproof}

\begin{lemma}
\label{DEXPR1primeminus}
For a DM-BIC with $Y_1\prec_o Y_2$ and a given $P\in\mathcal{P}'$, where $P$ factors as (\ref{newprop}), the set of DExPs of $\mathcal{R}_{(1),P}$ in Proposition \ref{Rprime} is given by $\Omega(\mathcal{R}_{(1),P})=\{E,F\}$, where $E$ and $F$ are given in Lemma \ref{DEXPR1minus}.
\end{lemma}

\begin{IEEEproof}
The proof is exactly the same as Case 1 in Appendix \ref{proofDEXPR1minus}.
\end{IEEEproof}

\begin{remark}
In Lemmas \ref{DEXPR2minus}-\ref{DEXPR1primeminus}, it is possible that for certain $P\in\mathcal{P}'$, some DExPs may coincide with one another. We do not explicitly handle this degenerate case.
\end{remark}

Since $\mathcal{R}_i=\bigcup_{P\in\mathcal{P}'}\mathcal{R}_{i,P}$, to find the set of DExPs of $\mathcal{R}_i$
%, denoted by $\Omega(\mathcal{R}_i)$,
we can first take the union of $\Omega(\mathcal{R}_{i,P})$ for all $P\in\mathcal{P}'$, and then remove the points which are dominated by some other point in the set. The set of DExPs of $\mathcal{R}_{(i)}$
%, denoted by $\Omega(\mathcal{R}_{(i)})$,
can be found similarly. Although Lemmas \ref{DEXPR2minus}-\ref{DEXPR1primeminus} suggest that $\Omega(\mathcal{R}_{(i),P})\subseteq\Omega(\mathcal{R}_{i,P})$ for a given $P\in\mathcal{P}'$, we will show in the following that the extra DExPs in $\Omega(\mathcal{R}_{i,P})$ will always be dominated by some DExPs in $\Omega(\mathcal{R}_{(i),\tilde{P}})$ evaluated for another $\tilde{P}\in\mathcal{P}'$.

\begin{IEEEproof}[Proof of Theorem \ref{theoremequi}]
We will first prove $\mathcal{R}_{2}=\mathcal{R}_{(2)}$ and then $\mathcal{R}_{1}=\mathcal{R}_{(1)}$. For a given $P\in\mathcal{P}'$, where $P$ factors as (\ref{newprop}), in order to simplify the notation, we use $P_{U_i=\phi}$, $i=1,2$, to denote the same probability distribution as $P$ except that $P_{U_i}(u_i)=\mathbbold{1}_{u_i,\phi}$, where $\mathbbold{1}_{u_i,\phi}$ is an indicator function defined in Introduction and $\phi$ is a constant. $P_{U_1=\phi,U_2=\phi}$ is defined similarly.

\noindent \textit{1. Proof of $\mathcal{R}_{2}=\mathcal{R}_{(2)}$}

From Lemma \ref{DEXPR2minus} and Lemma \ref{DEXPRpprimeminus}, for a given $P$, $\Omega(\mathcal{R}_{2,P})$ has two more DExPs $C,D$ than $\Omega(\mathcal{R}_{(2),P})$. However for $P_{U_1=\phi,U_2=\phi}$, we can show that point $A$, a DExP of $\Omega(\mathcal{R}_{(2),P})$, becomes
\begin{align*}
  A'=\left(\ I(X_1;Y_1),\ 0,\ I(X_2;Y_3)\ \right)
\end{align*}
and $C,D\leq A'$ due to $Y_1\succ_c Y_2$. This means that any extra DExP of $\Omega(\mathcal{R}_{2,P})$ for a given $P$ is always dominated by a DExP of $\Omega(\mathcal{R}_{(2),P})$ for $P_{U_1=\phi,U_2=\phi}$. Therefore $\mathcal{R}_{2}$ and $\mathcal{R}_{(2)}$ have identical DExPs. By Lemma \ref{DominateEx}, $\mathcal{R}_{2}=\mathcal{R}_{(2)}$.

\noindent \textit{2. Proof of $\mathcal{R}_{1}=\mathcal{R}_{(1)}$}

Although part 2 is more technical, the idea is essentially the same as part 1. We first consider the extra DExP $G$ of $\Omega(\mathcal{R}_{1,P})$. Given any $P$, consider another input distribution $P_{U_1=\phi}$, where $E,F$, DExPs of $\Omega(\mathcal{R}_{(1),P})$ become
\begin{align*}
    E' &= (\ 0,\ I(X_1;Y_2|U_2),\ \min\{ I(X_2;Y_3),\ I(U_2;Y_2)+I(X_2;Y_3|U_2) \} \ ), \\
    F' &= (\ 0,\ \min\{I(X_1;Y_2|U_2),\ [I(X_1,U_2;Y_2)-I(U_2;Y_3)]^+ \} ,\ I(X_2;Y_3|U_2)+ \\
    &\qquad \min\{ I(U_2;Y_3) ,\ I(X_1,U_2;Y_2) \} \ ).
\end{align*}
A region with DExPs given by $E'$ and $F'$ can be alternatively described as %(this can be verified by setting $U_1=\phi$ in $\mathcal{R}_{1}$)
\begin{align*}
    R_2&\leq I(X_1;Y_2|U_2)\\
    R_3&\leq I(X_2;Y_3)\\
    R_2+R_3&\leq I(X_1,U_2;Y_2)+I(X_2;Y_3|U_2)\\
    R_2,R_3&\geq 0, \ R_1=0.
\end{align*}
Since $I(U_1;Y_1)\leq I(U_1;Y_2)$ due to $Y_1\prec_o Y_2$, it can be shown that $G$ is contained in the above region. Hence $G$ is not a DExP after taking union of rate regions of all input distributions.

Next we will show that $H$, $I$ either reduce to or are dominated by some other DExPs once we consider all input distributions. Let us first focus on $H$.
\begin{enumerate}
\item If $I(U_2;Y_3)< I(U_2;Y_2|U_1)$, then
\begin{align*}
  G=(\ 0,\ I(U_1;Y_1)+I(X_1;Y_2|U_1,U_2),\ I(X_2;Y_3) \ ) \text{ and } H=G.
\end{align*}
\item If
\begin{align}
  I(U_2;Y_2|U_1) \leq I(U_2;Y_3) < I(X_1,U_2;Y_2|U_1) \label{dist},
\end{align}
we have
\begin{align*}
    F&=(\ I(U_1;Y_1),\ I(X_1,U_2;Y_2|U_1)-I(U_2;Y_3),\ I(X_2;Y_3) \ ),\\
    H&=(\ 0,\ I(U_1;Y_1)+ I(X_1,U_2;Y_2|U_1)-I(U_2;Y_3),\ I(X_2;Y_3) \ ).
\end{align*}
Notice that for any $P$ satisfying (\ref{dist}), $P_{U_1=\phi}$ also satisfies (\ref{dist}). Hence for $P_{U_1=\phi}$, $F$ becomes $F'=( 0,\ I(X_1,U_2;Y_2)-I(U_2;Y_3),\ I(X_2;Y_3) )$ and $H\leq F'$ due to $Y_1\prec_o Y_2$.
\item If $I(X_1,U_2;Y_2|U_1) \leq I(U_2;Y_3) \leq I(U_1;Y_1)+I(X_1,U_2;Y_2|U_1)$, we have
\begin{align*}
    H\leq G'=(\ 0,\ I(U_1;Y_1)+I(X_1;Y_2|U_1),\ I(X_2;Y_3) \ ),
\end{align*}
where $G'$ is $G$ for $P_{U_2=\phi}$.
\end{enumerate}

We now consider $I$.
\begin{enumerate}
\item If $I(U_2;Y_3)< I(U_2;Y_2|U_1)$, $I=E$.
\item If $I(U_2;Y_2|U_1)\leq I(U_2;Y_3)< I(X_1,U_2;Y_2|U_1)$, $I=F$.
\item If $I(X_1,U_2;Y_2|U_1)\leq I(U_2;Y_3)\leq I(U_1;Y_1)+I(X_1,U_2;Y_2|U_1)$, $I$ reduces to
\begin{align*}
    I=(\ I(U_1;Y_1)+ I(X_1,U_2;Y_2|U_1)-I(U_2;Y_3),\ 0,\ I(X_2;Y_3)\ ).
\end{align*}
For $P_{U_1=X_1,U_2=\phi}$, $E$ becomes $E'=( I(X_1;Y_1),\ 0,\ I(X_2;Y_3))$. Clearly, $I\leq E'$.
\end{enumerate}

Finally, we consider $J$. Given any $P$, for $P_{U_1=U_2=\phi}$, $E$ becomes
\begin{align*}
    E'=(\ 0,\ I(X_1;Y_2),\ I(X_2;Y_3) \ ).
\end{align*}
Clearly, $J<E'$ due to the condition $I(U_1;Y_1)+I(X_1,U_2;Y_2|U_1)< I(U_2;Y_3)$.

To summarize, even though for a given $P$, $\Omega(\mathcal{R}_{1,P})$ may be larger than $\Omega(\mathcal{R}_{(1),P})$, once we consider all $P\in\mathcal{P}'$, both regions $\mathcal{R}_{1}$ and $\mathcal{R}_{(1)}$ have exactly the same set of DExPs. By Lemma \ref{DominateEx}, $\mathcal{R}_{1}=\mathcal{R}_{(1)}$.
\end{IEEEproof}

%%%%%%%%%%%%%%%%%%%%%%%%%%%%%%%%%%%%%%%%%%%%____Capacity Regions_____%%%%%%%%%%%%%%%%%%%%%%%%%%%%%%%%%%%%%%%%%%%%
\section{Capacity Regions of DM-BICs under Partial-Order Broadcast Conditions and the Strong/Very Strong Interference Condition}
In Section IV, two simple achievable rate regions $\mathcal{R}_{(i)}$, $i=1,2$, are derived for DM-BICs with $Y_1\prec_o Y_2$ and $Y_1\succ_c Y_2$ respectively. Using these regions, in this section, capacity regions of DM-BICs with $Y_1\prec_o Y_2$ and $Y_1\succ_c Y_2$ will be established respectively for the strong and very strong interference conditions defined in the following.

\begin{definition}
\label{strongcondition}
In a DM-BIC, interference is said to be {\em strong} if for all $p(x_1)p(x_2)$, $I(X_2;Y_2|X_1)\geq I(X_2;Y_3)$.
\end{definition}
\begin{definition}
\label{verystrongcondition}
In a DM-BIC, interference is said to be {\em very strong} if for all $p(x_1)p(x_2)$, $I(X_2;Y_2)\geq I(X_2;Y_3)$.
\end{definition}

\begin{remark}
The intuition is similar as in a regular interference channel \cite{Costa}: For the strong interference, by conditioning on the intended signal, the interfered receiver sees a better channel than interfering user's own receiver. This suggests that the interfered receiver should be able to decode the interference along with its intended signal, by performing a joint decoding. Furthermore if interference is very strong, successive interference cancellation decoding suffices. Evidently very strong condition is stricter than the strong condition.
\end{remark}

\begin{theorem}
\label{capastrong}
The capacity region of a DM-BIC with $Y_1\prec_o Y_2$ and strong interference is the set of non-negative $(R_1,R_2,R_3)$ satisfying
\begin{align*}
    R_1&\leq I(U_1;Y_1)\\
    R_2&\leq I(X_1;Y_2|U_1,X_2)\\
    R_3&\leq I(X_2;Y_3)\\
    R_2+R_3&\leq I(X_1,X_2;Y_2|U_1),
\end{align*}
for some $P_{U_1X_1X_2}=P_{U_1}(u_1)P_{X_1|U_1}(x_1|u_1)P_{X_2}(x_2)$.
\end{theorem}

\begin{IEEEproof}
See Appendix \ref{proofcapastrong}.
\end{IEEEproof}

\begin{remark}
The capacity region takes two different forms. The one given in Theorem \ref{capastrong} is identical to $\mathcal{R}_{(1)}$ with $U_2=X_2$. An alternative form is given by $\mathcal{R}_{1}$ with $U_2=X_2$.
\end{remark}

When receiver 2 is interference-oblivious less noisy than receiver 1, for any sensible coding scheme $X_1$ should always be decodable at receiver 2 (otherwise, none of the broadcast receivers can do so). Hence the strong condition, originated from a regular interference channel, naturally carries over to a DM-BIC with $Y_1\prec_o Y_2$. However, this is not the case for a DM-BIC with $Y_1\succ_c Y_2$, which will be discussed next.

\begin{theorem}
\label{capaverystrong}
The capacity region of a DM-BIC with $Y_1\succ_c Y_2$ and very strong interference is the set of non-negative $(R_1,R_2,R_3)$ satisfying
\begin{align*}
    R_1&\leq I(X_1;Y_1|U_1)\\
    R_2&\leq I(U_1;Y_2|X_2)\\
    R_3&\leq I(X_2;Y_3),
\end{align*}
for some $P_{U_1X_1X_2}=P_{U_1}(u_1)P_{X_1|U_1}(x_1|u_1)P_{X_2}(x_2)$.
\end{theorem}

\begin{IEEEproof}
The achievability follows those for $\mathcal{R}_{(2)}$ and $\mathcal{R}_{2}$ with $U_2=X_2$. The converse proof is straightforward and hence omitted.
\end{IEEEproof}

\begin{remark}
Similar to Theorem \ref{capastrong}, the capacity region takes two forms, $\mathcal{R}_{(2)}$ with $U_2=X_2$ and $\mathcal{R}_{2}$ with $U_2=X_2$.
\end{remark}

It is not difficult to see that the strong interference condition in Definition \ref{strongcondition} does not fit well for a DM-BIC with $Y_1\succ_c Y_2$. The reason is that the conditioning of $X_1$ in Definition \ref{strongcondition} seems to imply that $X_1$ is the intended signal for receiver 2, i.e. $X_1$ is always decodable at receiver 2. Then by $Y_1\succ_c Y_2$, receiver 1 can decode it as well. Hence the two receivers will always decode the same set of messages, which clearly does not represent the most general case. In fact, we claim that the strong but not very strong interference condition does not exist for a DM-BIC with $Y_1\succ_c Y_2$. The argument is as follows.

As in the strong interference condition for a regular interference channel \cite{Costa}, the problem is to figure out what is the intended signal for receiver 2, rather than simply conditioning on $X_1$. Once we have found such a signal, we can mimic the strong condition in Definition \ref{strongcondition}, with modification of conditioning on that signal instead of $X_1$.
%Suppose there exists some strong condition, then interference $X_2$ should be decoded at receiver 2. Under this restriction, we have an upper bound $n(R_2+R_3-\epsilon_n)\leq I(W_2,W_3;Y_2^n)$. Along with other straightforward upper bounds, by the same technique that we used above to prove Theorem \ref{capastrong}, we can show that $\mathcal{R}_{(2)}$ with $U_2=X_2$ is the capacity region. This implies that if there exists some strong condition, then superposition coding with cloud center $U_1$ carrying receiver 2's message is capacity achieving. Hence without loss of generality, we can view the cloud center $U_1$ as the intended signal for receiver 2, which in return gives us the strong condition
Following the converse (similar to the converse proof of Theorem \ref{capastrong}) with the condition that interference signal $X_2$ should be decodable at receiver 2 (as in the usual strong interference condition), we would get $I(X_2;Y_2|U_1)\geq I(X_2;Y_3)$, for all $p(u_1)p(x_1|u_1)p(x_2)$ such that $U_1\rightarrow (X_1,X_2)\rightarrow (Y_2,Y_3)$ form a Markov chain. However, since this holds for all $U_1$, it would imply the very strong interference condition ($U_1=\phi$). Therefore for a DM-BIC with $Y_1\succ_c Y_2$, a meaningful strong but not very strong interference condition does not exists.

%%%%%%%%%%%%%%%%%%%%%%%%%%%%%%%%%%%%%%%%%%%%%%%%____Gaussian____%%%%%%%%%%%%%%%%%%%%%%%%%%%%%%%%%%%%%%%%%%%%
\section{Gaussian Broadcast Interference Channel}
In this section, we extend the DM-BIC results to the Gaussian case. To guide our discussions for the GBIC in (\ref{channel}), we will divide the channel parameters into three regimes according to receiver 2's broadcast link strength: $a\geq 1+bP_2$, $1< a< 1+bP_2$ and $0\leq a\leq 1$. Our focus will be on the first two cases, where the derived achievable rate regions and outer bounds will be shown to coincide under various conditions. For $0\leq a\leq 1$, most of the results can be deduced from \cite{Shang II}, but we include this case for completeness and we also provide simpler characterizations of the inner and outer bounds. At the end of this section, we present some numerical results to demonstrate the tightness of our bounds.

%____________________________________________________________________________________________________________________
\subsection{GBIC with $a\geq 1+bP_2$}
Assuming $X_2$ is Gaussian distributed, it can be seen that, when $a\geq 1+bP_2$, $I(U_1;Y_1)\leq I(U_1;Y_2)$ for all $p(u_1,x_1)$ such that $U_1\rightarrow (X_1,X_2)\rightarrow (Y_1,Y_2)$ form a Markov chain. Hence we obtain an inner bound for a GBIC by specializing the inner bound for a DM-BIC with $Y_1\prec_o Y_2$ with Gaussian inputs.

\begin{corollary}
\label{coro}
For some $\alpha,\gamma \in[0,1]$, let $\mathcal{S}_1(\alpha,\gamma)$ denote the set of non-negative $(R_1,R_2,R_3)$ satisfying
\begin{align*}
    R_1&\leq \mathcal{C}\left(\tfrac{\bar{\alpha} P_1}{1+\alpha P_1}\right)\\
    R_2&\leq  \mathcal{C}\left(\tfrac{a\alpha P_1}{1+b\bar{\gamma}P_2}\right)\\
    R_3&\leq \mathcal{C}(P_2)\\
    R_2+R_3&\leq \mathcal{C}\left(\tfrac{a\alpha P_1+b\gamma P_2}{1+b\bar{\gamma}P_2}\right) + \mathcal{C}(\bar{\gamma}P_2).
\end{align*}
Then $\mathcal{S}_1=\bigcup_{\alpha,\gamma\in[0,1]}\mathcal{S}_1(\alpha,\gamma)$ is an achievable rate region of the GBIC in (\ref{channel}) with $a\geq 1+bP_2$.
\end{corollary}

\begin{IEEEproof}
We evaluate $\mathcal{R}_{(1)}$ given in Proposition \ref{Rprime} for a DM-BIC with $Y_1\prec_o Y_2$. Let $U_1=\sqrt{\bar{\alpha}}X_{11}$ and $X_1=\sqrt{\bar{\alpha}}X_{11}+\sqrt{\alpha}X_{12}$ for some $\alpha\in[0,1]$, where $X_{11},X_{12}\sim \mathcal{N}(0,P_1)$ and are independent. Similarly $U_2=\sqrt{\gamma}X_{2c}$ and $X_2=\sqrt{\gamma}X_{2c}+\sqrt{\bar{\gamma}}X_{2p}$, for some $\gamma\in[0,1]$, where $X_{2c},X_{2p}\sim \mathcal{N}(0,P_2)$ and are independent. Evaluating $\mathcal{R}_{(1)}$ with the above random variables, we obtain $\mathcal{S}_1$.
\end{IEEEproof}

\begin{remark}
In the achievable scheme, the broadcast transmitter employs superposition coding, with $\alpha$ representing the power distribution between the signals for receiver 1 and 2. The interference transmitter employs rate splitting, with $\gamma$ representing the power distribution between the common and private signals. In fact, with this coding scheme, one can directly obtain an inner bound without resorting to $\mathcal{R}_{(1)}$. However, we note that condition $Y_1\prec_o Y_2$ is used to remove redundant constraints in the derivation of $\mathcal{R}_{(1)}$. This leads to a concise representation of the inner bound as shown in Corollary \ref{coro}.
\end{remark}

Unlike the achievability part, the converse result given by Theorem \ref{capastrong} for a DM-BIC with $Y_1\prec_o Y_2$ cannot be specialized to the GBIC with $a\geq 1+bP_2$. This is because the GBIC with $a\geq 1+bP_2$ does not satisfy the condition of $Y_1\prec_o Y_2$ for an arbitrary distribution of $X_2$. Next, we derive a general outer bound for the GBIC with $a\geq 1+bP_2$ using EPI. In order to invoke EPI to relate broadcast rates $R_1$ and $R_2$, we first need to quantify the loss of optimality when fixing $X_2$ to be Gaussian. To achieve this, we use a result from \cite{Zamir}, which says that the Gaussian input incurs no more than half-bit loss compared to the optimal distribution for an arbitrary additive noise channel.

\begin{theorem}
\label{theo:outerbound2}
Let $\mathcal{O}_1(\alpha)$ denote the set of non-negative $(R_1,R_2,R_3)$ satisfying the following
\begin{align}
    R_1&\leq \mathcal{C}\left(\tfrac{\bar{\alpha}P_1}{1+\alpha P_1}\right) \label{EQ3}\\
    R_2&\leq \mathcal{C}(a\alpha P_1+bP_2) - \xi(b) + 0.5 \label{EQ}\\
    R_2&\leq \mathcal{C}(aP_1+bP_2) - \xi(b) \label{EQ2}\\
    R_2&\leq \mathcal{C}(a\alpha P_1) \label{EQ4}\\
    R_3&\leq \mathcal{C}(P_2),\label{EQ5}
\end{align}
for some $\alpha\in[0,1]$, where
\begin{align}
    \xi(x)\triangleq\begin{cases}
    \mathcal{C}(x(2^{2R_3}-1)),\quad x<1\\
    R_3,\quad x\geq 1
    \end{cases}.\label{xi}
\end{align}
Then $\mathcal{O}_1=\bigcup_{\alpha\in[0,1]}\mathcal{O}_1(\alpha)$ is an outer bound on the capacity region of the GBIC in (\ref{channel}) with $a\geq 1+bP_2$.
\end{theorem}

\begin{IEEEproof}
See Appendix \ref{prooftheo:outerbound2}.
\end{IEEEproof}

When interference is strong but not very strong, i.e. $1\leq b< 1+aP_1$, a looser outer bound can be derived by dropping (\ref{EQ2}) from $\mathcal{O}_1(\alpha)$. This new outer bound differs from the inner bound $\mathcal{S}_1(\alpha,\gamma)$ with $\gamma=1$ only by a constant $0.5$ in inequality (\ref{EQ}). Hence having receiver $Y_2$ decoding interference as a whole is approximately capacity achieving.

\begin{corollary}
\label{coro:halfbit}
The inner bound $\mathcal{S}_1$ given in Corollary \ref{coro}, is within $0.5$ bits to the capacity region of the GBIC in (\ref{channel}) with $a\geq 1+bP_2$ and $1\leq b< 1+aP_1$.
\end{corollary}

It is also straightforward to obtain the capacity region when interference is very strong, i.e. $b\geq 1+aP_1$.
\begin{corollary}
\label{coroverystrong}
The capacity region of the GBIC in (\ref{channel}) with $a\geq 1+bP_2$ and $b\geq 1+aP_1$ is given by $\{(R_1,R_2,R_3):0\leq R_1\leq \mathcal{C}(\frac{\bar{\alpha} P_1}{1+\alpha P_1}), 0\leq R_2\leq \mathcal{C}(a\alpha P_1), 0\leq R_3\leq \mathcal{C}(P_2) \}$.
\end{corollary}

In the following, we focus on sum rate/capacity analysis.
\begin{proposition}
The largest sum rate of the achievable rate region given in Corollary \ref{coro} is
\begin{align}
    R_s=R_1+R_2+R_3=\begin{cases}
        \mathcal{C}\left(\tfrac{aP_1}{1+bP_2}\right) + \mathcal{C}(P_2),\quad \textrm{if }0\leq b< 1\\
        \min\{\mathcal{C}(aP_1+bP_2),\mathcal{C}(aP_1)+\mathcal{C}(P_2)\}, \quad \textrm{if }b\geq 1
    \end{cases} \label{Rsum}
\end{align}
\end{proposition}

\begin{IEEEproof}
$R_s$ is achieved by setting $\alpha = 1$, $\gamma = 0$ if $b< 1$ and $\alpha = 1$, $\gamma = 1$ if $b\geq 1$. The sum rate optimality of setting $\alpha=1$ among all achievable schemes in Corollary \ref{coro} follows from the fact that $Y_1$ is stochastically degraded with respect to $Y_2$ when $X_1$ and $X_2$ are Gaussian random variables and setting $\alpha=1$ maximizes $R_1+R_2$ irrespective of interference. With $\alpha=1$, the channel becomes a Gaussian Z interference channel (GZIC). The sum rate optimality of setting $\gamma=0$ if $b< 1$ and $\gamma=1$ if $b\geq 1$ comes from the sum capacity result of a GZIC \cite{Sason}.
\end{IEEEproof}

\begin{theorem}
\label{theo:halfbitsumcap}
The sum rate $R_s$ given in (\ref{Rsum}) is within $0.5$ bits to the sum capacity $C_s$ of the GBIC in (\ref{channel}) with $a\geq 1+bP_2$, i.e. $C_s\leq R_s+0.5$.
\end{theorem}

\begin{IEEEproof}
See Appendix \ref{prooftheo:halfbitsumcap}.
\end{IEEEproof}

%____________________________________________________________________________________________________________________
\subsection{GBIC with $1< a< 1+bP_2$}
With $1< a< 1+bP_2$, there is no immediate ordering of the decodability between the two broadcast receivers given the interference from the point-to-point transmission. We consider two superposition inner bounds, where the signal for each of the broadcast receivers serves as the cloud center respectively, and then take convex hull of the two regions.

\begin{corollary}
\label{coro:s3}
For some $\alpha_1,\alpha_2,\gamma_1,\gamma_2 \in[0,1]$, let $\mathcal{S}_2(\alpha,\gamma)$ denote the set of non-negative $(R_1,R_2,R_3)$ satisfying
\begin{align}
    R_1&\leq \mathcal{C}\left(\tfrac{\bar{\alpha}_1P_1}{1+\alpha_1P_1}\right)\notag\\
    R_3&\leq \mathcal{C}(P_2)\label{rs2b}\\
    R_1+R_2&\leq \mathcal{C}\left(\tfrac{aP_1}{1+b\bar{\gamma}_1P_2}\right)\label{rs2d}\\
    R_1+R_2&\leq \mathcal{C}\left(\tfrac{\bar{\alpha}_1P_1}{1+\alpha_1P_1}\right) + \mathcal{C}\left(\tfrac{a\alpha_1P_1}{1+b\bar{\gamma}_1P_2}\right)\label{rs2e}\\
    R_1+R_2+R_3&\leq \mathcal{C}\left(\tfrac{aP_1+b\gamma_1P_2}{1+b\bar{\gamma}_1P_2}\right) + \mathcal{C}(\bar{\gamma}_1 P_2)\label{rs2f}\\
    R_1+R_2+R_3&\leq \mathcal{C}\left(\tfrac{\bar{\alpha}_1P_1}{1+\alpha_1P_1}\right) + \mathcal{C}\left(\tfrac{a\alpha_1P_1+b\gamma_1 P_2}{1+b\bar{\gamma}_1P_2}\right) + \mathcal{C}(\bar{\gamma}_1 P_2),\label{rs2g}
\end{align}
and let $\mathcal{S}_3(\alpha,\gamma)$ denote the set of non-negative $(R_1,R_2,R_3)$ satisfying
\begin{align}
    R_2&\leq \mathcal{C}\left(\tfrac{a\alpha_2 P_1}{1+a\bar{\alpha}_2P_1+b\bar{\gamma}_2 P_2}\right)\notag\\
    R_3&\leq \mathcal{C}(P_2)\label{rs1b}\\
    R_1+R_2&\leq \mathcal{C}(\bar{\alpha}_2 P_1) + \mathcal{C}\left(\tfrac{a\alpha_2 P_1}{1+a\bar{\alpha}_2P_1+b\bar{\gamma}_2 P_2}  \right)\label{rs1e}\\
    R_1+R_2&\leq \mathcal{C}(P_1)\label{rs1d}\\
    R_2+R_3&\leq \mathcal{C}\left(\tfrac{a\alpha_2 P_1+b\gamma_2 P_2}{1+a\bar{\alpha}_2P_1+b\bar{\gamma}_2 P_2}\right) + \mathcal{C}(\bar{\gamma}_2 P_2)\notag\\
    R_1+R_2+R_3&\leq \mathcal{C}(\bar{\alpha}_2 P_1) + \mathcal{C}\left(\tfrac{a\alpha_2 P_1+b\gamma_2 P_2}{1+a\bar{\alpha}_2P_1+b\bar{\gamma}_2 P_2}\right) + \mathcal{C}(\bar{\gamma}_2 P_2).\label{rs1g}
\end{align}
We further let $\mathcal{S}_2=\bigcup_{\alpha_1,\gamma_1\in[0,1]}\mathcal{S}_2(\alpha,\gamma)$ and $\mathcal{S}_3=\bigcup_{\alpha_1,\gamma_1\in[0,1]}\mathcal{S}_3(\alpha,\gamma)$. Then the convex hull of $\mathcal{S}_2\bigcup\mathcal{S}_3$ is an achievable rate region of the GBIC in (\ref{channel}) with $1< a< 1+bP_2$.
\end{corollary}

\begin{IEEEproof}
See Appendix \ref{proofcoro:s3}.
\end{IEEEproof}

\begin{proposition}
\label{prop:sumrate}
The sum rate of the inner bound in Corollary \ref{coro:s3} is given by $R_s=\max\{R_{s,1},R_{s,2}\}$, where
\begin{align*}
    R_{s,1}&=\max_{\alpha_1,\gamma_1\in[0,1]}\min\{\textrm{rhs}(\ref{rs2b})+\textrm{rhs}(\ref{rs2d}),\ \textrm{rhs}(\ref{rs2b})+\textrm{rhs}(\ref{rs2e}),\ \textrm{rhs}(\ref{rs2f}),\ \textrm{rhs}(\ref{rs2g})\},\\
    R_{s,2}&=\mathcal{C}(P_1)+\mathcal{C}(P_2),
\end{align*}
and $\textrm{rhs}(x)$ denotes the right-hand side of inequality $(x)$.
\end{proposition}

\begin{IEEEproof}
It is straightforward to show that the sum rates of $\mathcal{S}_2(\alpha,\gamma)$ is given by $\min\{\textrm{rhs}(\ref{rs2b})+\textrm{rhs}(\ref{rs2d}), \textrm{rhs}(\ref{rs2b})+\textrm{rhs}(\ref{rs2e}), \textrm{rhs}(\ref{rs2f}), \textrm{rhs}(\ref{rs2g})\}$. Similarly the sum rate of $\mathcal{S}_3(\alpha,\gamma)$ is given by $\min\{\textrm{rhs}(\ref{rs1b})+\textrm{rhs}(\ref{rs1e}), \textrm{rhs}(\ref{rs1b})+\textrm{rhs}(\ref{rs1d}), \textrm{rhs}(\ref{rs1g})\}$. Since $\textrm{rhs}(\ref{rs1b}) +\textrm{rhs}(\ref{rs1d})$ is indeed achievable with $\alpha_2=\gamma_2=0$, $R_{s,2}=\mathcal{C}(P_1) + \mathcal{C}(P_2)$.
\end{IEEEproof}

\begin{theorem}
\label{theo:outerbound3}
Let $\mathcal{O}_2(\alpha)$ denote the set of non-negative $(R_1,R_2,R_3)$ satisfying the following
\begin{align}
    R_1&\leq \mathcal{C}\left(\tfrac{\bar{\alpha}P_1}{1+\alpha P_1} \right) \notag\\
    R_2&\leq \mathcal{C}(a\alpha P_1) \notag\\
    R_2&\leq \mathcal{C}(aP_1+bP_2) - \xi(b) \label{boundeq}\\
    R_3&\leq \mathcal{C}(P_2),\notag
\end{align}
where $\alpha\in[0,1]$ and $\xi(\cdot)$ is defined in (\ref{xi}). Then $\mathcal{O}_2=\bigcup_{\alpha\in[0,1]}\mathcal{O}_2(\alpha)$ is an outer bound on the capacity region of the GBIC in (\ref{channel}) with $1< a< 1+bP_2$.
\end{theorem}

\begin{IEEEproof}
The constraint (\ref{boundeq}) on $R_2$ is the same as (\ref{EQ2}) and the argument follows similarly. The remaining constraints are obtained by removing the interfering link and invoking the capacity results of a GBC and a point-to-point channel.
\end{IEEEproof}

%____________________________________________________________________________________________________________________
\subsection{GBIC with $0\leq a\leq 1$}
Channel condition $0\leq a\leq 1$ can be viewed as the Gaussian counterpart of interference-cognizant less noisy $Y_1\succ_c Y_2$. To see this, we note that, conditioned on $X_2$, $Y_2$ is stochastically degraded \cite{Cover} with respect to $Y_1$, which implies $I(U_1;Y_1)\geq I(U_1;Y_2|X_2)$ for all $p(u_1,x_1)p(x_2)$ such that $U_1\rightarrow (X_1,X_2)\rightarrow (Y_1,Y_2)$ form a Markov chain. The following inner bound for the GBIC is a direct consequence of this relation.

\begin{corollary}
\label{coro:s1}
For some $\alpha,\gamma \in[0,1]$, let $\mathcal{S}_4(\alpha,\gamma)$ denote the set of non-negative $(R_1,R_2,R_3)$ satisfying
\begin{align*}
    R_1&\leq \mathcal{C}(\bar{\alpha}P_1)\\
    R_2&\leq \mathcal{C}\left(\tfrac{a\alpha P_1}{1+a\bar{\alpha}P_1+b\bar{\gamma}P_2}\right)\\
    R_3&\leq \mathcal{C}(P_2)\\
    R_2+R_3&\leq \mathcal{C}\left(\tfrac{a\alpha P_1+b\gamma P_2}{1+a\bar{\alpha}P_1+b\bar{\gamma}P_2}\right) + \mathcal{C}(\bar{\gamma}P_2).
\end{align*}
Then $\mathcal{S}_4=\bigcup_{\alpha,\gamma\in[0,1]}\mathcal{S}_4(\alpha,\gamma)$ is an achievable rate region for the GBIC in (\ref{channel}) with $0\leq a\leq 1$.
\end{corollary}

\begin{IEEEproof}
Since $0\leq a\leq 1$ corresponds to $Y_1\succ_c Y_2$, we evaluate the inner bound $\mathcal{R}_{(2)}$. Specifically, we let $U_1=\sqrt{\alpha}X_{12}$ and $X_1=\sqrt{\bar{\alpha}}X_{11}+\sqrt{\alpha}X_{12}$ for some $\alpha\in[0,1]$, where $X_{11},X_{12}\sim \mathcal{N}(0,P_1)$ are independent. Similarly $U_2=\sqrt{\gamma}X_{2c}$ and $X_2=\sqrt{\gamma}X_{2c}+\sqrt{\bar{\gamma}}X_{2p}$, for some $\gamma\in[0,1]$, where $X_{2c},X_{2p}\sim \mathcal{N}(0,P_2)$ are independent. Evaluating $\mathcal{R}_{(2)}$ with the above random variables, we obtain $\mathcal{S}_4$.
\end{IEEEproof}

In Section V, it was observed that the usual strong but not very strong interference condition in an IC does not carry over to a DM-BIC with $Y_1\succ_c Y_2$, similarly to the GBIC with $0\leq a\leq 1$. In a regular IC, the strong interference condition is obtained by comparing the interference-to-noise-ratio (INR) at the interfered receiver and the signal-to-noise-ratio at the desired receiver. However in the GBIC, the dynamic of the trade-off between $R_1$ and $R_2$ for the broadcast part causes the effective noise level at receiver 2 to vary, amounting to a varying INR. Therefore the usual notion of strong interference does not apply to the GBIC. However, it is easy to see that the very strong interference condition, i.e. $b\geq 1+aP_1$, still holds for the GBIC, leading to the following corollary.

\begin{corollary}
The capacity region of the GBIC in (\ref{channel}) with $0\leq a\leq 1$ and $b\geq 1+aP_1$ is given by $\{(R_1,R_2,R_3):0\leq R_1\leq \mathcal{C}(\bar{\alpha}P_1), 0\leq R_2\leq \mathcal{C}(\frac{a\alpha P_1}{1+a\bar{\alpha} P_1}), 0\leq R_3\leq \mathcal{C}(P_2) \}$.
\end{corollary}

For the general case, in particular when $b< 1+aP_1$, we present an EPI-based outer bound in the following theorem, which is a special case of the outer bound in \cite{Shang II} albeit has a simpler representation.

\begin{theorem}
\label{theo:outerbound1}
Let $\mathcal{O}_4(\alpha)$ denote the set of non-negative $(R_1,R_2,R_3)$ satisfying
\begin{align}
    R_1&\leq \mathcal{C}(\bar{\alpha} P_1)\notag\\
    R_2&\leq \mathcal{C}\left( \tfrac{a\alpha P_1+bP_2+(1-a)(1-2^{2\xi(\frac{b}{1-a})})}{a+a\bar{\alpha} P_1+(1-a)2^{2\xi(\frac{b}{1-a})}} \right)\label{boundsR2}\\
    R_2&\leq \mathcal{C}\left(\tfrac{a\alpha P_1}{1+a\bar{\alpha} P_1} \right) \label{boundsR22}\\
    R_3&\leq \mathcal{C}(P_2),\notag
\end{align}
where $\alpha\in[0,1]$ and $\xi(\cdot)$ is defined in (\ref{xi}). Then $\mathcal{O}_4=\bigcup_{\alpha\in[0,1]}\mathcal{O}_4(\alpha)$ is an outer bound on the capacity region of the GBIC in (\ref{channel}) with $0\leq a\leq 1$.
\end{theorem}

\begin{IEEEproof}
The proof is similar to that of Theorem \ref{theo:outerbound2} and is omitted.
%See Appendix \ref{prooftheo:outerbound1}.
\end{IEEEproof}

%Equipped with $\mathcal{O}_4$, we are able to determine the capacity region boundary partially when $b< 1+aP_1$. For any achievable rate triple $(R_1,R_2,R_3)$ in Corollary \ref{coro:s1}, there exists a $\beta\in[0,1]$ such that $R_3=\mathcal{C}(\beta P_2)$. In the following, we show that when $\beta$ is sufficiently small: $\beta\leq\frac{\min\{1,b\}}{1+aP_1}$, or $\beta=1$, the achievable scheme in Corollary \ref{coro:s1} is capacity achieving.

\begin{theorem}
\label{theo:partial}
The following rate triples are on the boundary of the capacity region of the GBIC in (\ref{channel}) with $0\leq a\leq 1$:
{\small
\begin{align}
\left\{
\begin{aligned}
&(R_1,R_2,R_3):\ R_1=\mathcal{C}(\bar{\alpha}P_1),\ R_2=\mathcal{C}\left(\tfrac{a\alpha P_1}{1+a\bar{ \alpha}P_1}\right),\\
&R_3=\mathcal{C}(\beta P_2),\ \text{for some }\alpha\in[0,1],\ \beta\in[0,\min\{1,\tfrac{b}{1+aP_1}\}]
\end{aligned}
\right\}\label{partial1}
\end{align}
}and if $a+b\leq 1$
{\small
\begin{align}
\left\{
\begin{aligned}
&(R_1,R_2,R_3):\ R_1=\mathcal{C}(\bar{\alpha}P_1),\ R_2=\mathcal{C}\left(\tfrac{a\alpha P_1}{1+a\bar{ \alpha}P_1+bP_2}\right),\\
&R_3=\mathcal{C}(P_2),\ \text{for some }\alpha\in[0,1]
\end{aligned}
\right\}\label{partial2}
\end{align}
}
\end{theorem}

\begin{IEEEproof}
To get (\ref{partial1}), it is easy to see that when $\beta\in[0,\min\{1,\tfrac{b}{1+aP_1}\}]$, receiver 2 is able to decode the interference signal by treating the broadcast signal as noise. This disassociates the broadcast component from the interference component of the channel. Hence $(R_1,R_2)$ are given by the GBC capacity region and $R_3$ is given by the point-to-point rate $\mathcal{C}(\beta P_2)$. To get (\ref{partial2}), we can show that $\mathcal{S}_4(\alpha,\gamma)$ with $\gamma = 0$ coincides with $\mathcal{O}_4(\alpha)$ when $a+b\leq 1$.
%See Appendix \ref{prooftheo:partial}.
\end{IEEEproof}

\begin{corollary}
The sum capacity of the GBIC in (\ref{channel}) with $0\leq a\leq 1$ is $\mathcal{C}(P_1) + \mathcal{C}(P_2)$.
\end{corollary}

\begin{IEEEproof}
$\mathcal{C}(P_1) + \mathcal{C}(P_2)$ is the sum rate of $\mathcal{S}_4(0,0)$ in Corollary \ref{coro:s1}. The converse follows by realizing $R_1+R_2\leq \mathcal{C}(P_1)$ when the interference link is removed and the obvious upper bound $R_3\leq \mathcal{C}(P_2)$.
\end{IEEEproof}

%____________________________________________________________________________________________________________________
\subsection{Numerical Results}
\begin{figure}[htb]
  \centering
  \psfrag{beta0.9}[c][c][0.7][0]{$\beta=0.9$}
  \psfrag{beta0.4}[l][l][0.7][0]{$\beta=0.4$}
  \psfrag{beta0.1}[l][l][0.7][0]{$\beta=0.1$}
  \psfrag{R1}[c][c][0.8][0]{$R_1$}
  \psfrag{R2}[c][c][0.8][-90]{$R_2$}
  \psfrag{R3}{$R_3=\mathcal{C}(\beta P_2)$}
  \includegraphics[width=80mm]{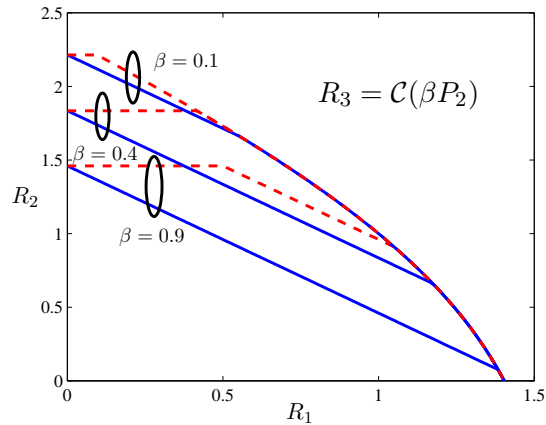}
  \caption{Inner (solid line) and outer bounds (dashed line) for the GBIC in (\ref{channel}) with $P_1=6$, $P_2=3$, $a=4$, $b=1$}
  \label{fig:Region2}
\end{figure}

For $a\geq 1+bP_2$, inner bound $\mathcal{S}_1$ and outer bound $\mathcal{O}_1$ are plotted in Fig. \ref{fig:Region2}, where $R_3=\mathcal{C}(\beta P_2)$ is fixed at different values to investigate the trade-off between $R_1$ and $R_2$. We observe that for all values of $\beta$, the inner bound is always within 0.5 bits to the outer bound, confirming Corollary \ref{coro:halfbit}.

\begin{figure}[htb]
  \centering
  \psfrag{Rs1}[c][c][0.7][0]{$\, R_{s,1}$}
  \psfrag{Rs2}[c][c][0.7][0]{$\, R_{s,2}$}
  \psfrag{Rs}[c][c][0.7][0]{$R_s$}
  \psfrag{Ro}[c][c][0.7][0]{$R_o$}
  \psfrag{a}{$a$}
  \includegraphics[width=80mm]{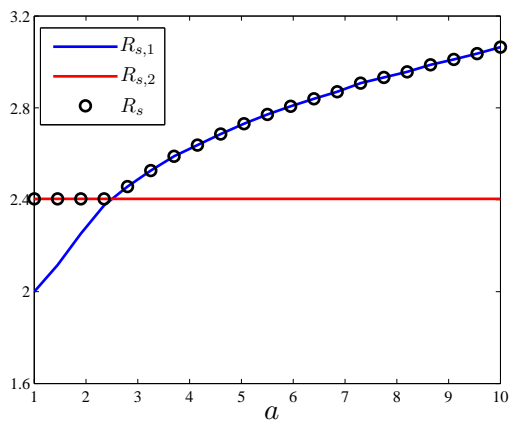}
  \caption{Sum rate for the GBIC in (\ref{channel}) with $P_1=6$, $P_2=3$, $b=3$, $1< a< 1+bP_2$}
  \label{fig:sumrate}
\end{figure}

For $1< a< 1+bP_2$, the sum rate from Proposition \ref{prop:sumrate} is plotted in Fig. \ref{fig:sumrate}, where sum rate upper bound $R_o$ is obtained from $\mathcal{O}_2$ in Theorem \ref{theo:outerbound3}. We observe that depending on the strength of broadcast link $a$, one of the two regions, $\mathcal{S}_2$ and $\mathcal{S}_3$, gives rise to $R_s$, ensuring the necessity of the convex hull operation in the achievable scheme given by Corollary \ref{coro:s3}.

\begin{figure}[htb]
  \centering
  \psfrag{beta0.9}[c][c][0.7][0]{$\beta=1$}
  \psfrag{beta0.6}[c][c][0.7][0]{$\beta=0.6$}
  \psfrag{beta0.3}[r][r][0.7][0]{$\beta=0.3$}
  \psfrag{beta0.1}[l][l][0.7][0]{$\beta=0.1$}
  \psfrag{R1}[c][c][0.8][0]{$R_1$}
  \psfrag{R2}[c][c][0.8][-90]{$R_2$}
  \psfrag{R3}{$R_3=\mathcal{C}(\beta P_2)$}
  \includegraphics[width=80mm]{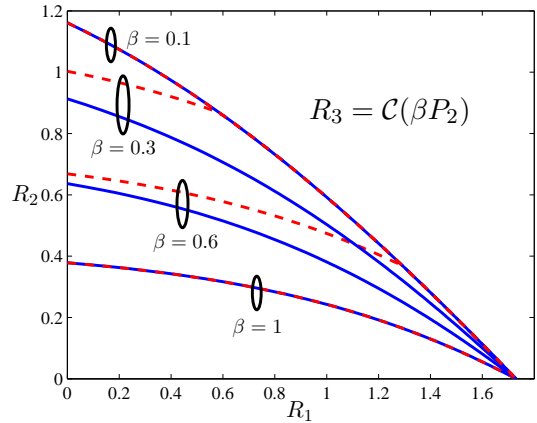}
  \caption{Inner (solid line) and outer bounds (dashed line) for the GBIC in (\ref{channel}) with $P_1=10$, $P_2=8$, $a=0.4$, $b=0.6$}
  \label{fig:Region1}
\end{figure}

For $0\leq a\leq 1$, inner bound $\mathcal{S}_4$ and outer bound $\mathcal{O}_4$ are plotted in Fig. \ref{fig:Region1}. The result agrees with Theorem \ref{theo:partial}. We observe that these bounds are tight when the interference rate is either small or equal to $\mathcal{C}(P_2)$.

%%%%%%%%%%%%%%%%%%%%%%%%%%%%%%%%%%%%%%%%%%%%%%%%____Conclusion_____%%%%%%%%%%%%%%%%%%%%%%%%%%%%%%%%%%%%%%%%%%%%
\section{Conclusion}
In this paper, we have proposed a coding scheme that combines message splitting, superposition coding and binning for a general DM-BIC. We have then specialized the obtained achievable rate region to DM-BICs under two partial-order broadcast conditions: interference-oblivious less noisy and interference-cognizant less noisy. By carefully inspecting the dominant extreme points, we have shown the specialized rate region to be equivalent to a rate region based on a simpler scheme, where the broadcast transmitter uses only superposition coding. For the interference-oblivious less noisy DM-BIC, if interference is strong, the capacity region is given by the aforementioned two equivalent rate regions. For the interference-cognizant less noisy DM-BIC, we have argued that the strong but not very strong interference condition does not exist and we have obtained the capacity region for very strong interference. For a GBIC, we have divided channel parameters into three regimes and obtained achievable rate regions and capacity region outer bounds for each of these regimes. Noticeably, the outer bounds extend previously known bounds in the literature by combing bounding techniques for a GBC and a GIC. We have subsequently determined capacity regions and sum capacity either exactly or approximately (within a half-bit) under various conditions.

%%%%%%%%%%%%%%%%%%%%%%%%%%%%%%%%%%%%%%%%%%%%%%%____Appendix_____%%%%%%%%%%%%%%%%%%%%%%%%%%%%%%%%%%%%%%%%%%%%%%%%%
\appendices
%------------------------------------------------------------------------------------------------------------------
\section{Proof of Lemma \ref{R3hat}}
\label{ProofLemmaR3hat}
\begin{IEEEproof}
In the following, we prove the achievability of $\hat{\mathcal{R}}_P$ for a fixed $P\in\hat{\mathcal{P}}$. %By considering all possible $P\in\hat{\mathcal{P}}$, we obtain an achievable rate region $\hat{\mathcal{R}}$.

\textit{Codebook generation}:

We first generate time sharing sequence $q^n$ with each symbol i.i.d according to $p_Q$. Receiver 1's message is split into common message $m_1$, to be decoded by both receiver 1 and receiver 2, and private message $i$, to be decoded only by receiver 1. Similarly receiver 2's message is split into common message $m_2$, to be decoded by both receiver 1 and receiver 2, and private message $j$, to be decoded only by receiver 2. We generate $2^{n(R_{1c}+R_{2c})}$ independent codewords $u_1^n(m_1,m_2)$ with each symbol i.i.d according to $p_{U_1|Q}$, $m_1\in\{1,...,2^{nR_{1c}}\}$, $ m_2\in\{1,...,2^{nR_{2c}}\}$. For each $u_1^n(m_1,m_2)$, we generate $2^{n(R_{1p}+R_1')}$ conditionally independent codewords $v_1^n(m_1,m_2,i,i')$ with each symbol i.i.d according to $p_{V_1|U_1,Q}$, $i\in\{1,...,2^{nR_{1p}}\}$, $ i'\in\{1,...,2^{nR_1'}\}$. Similarly for each $u_1^n(m_1,m_2)$, we generate $2^{n(R_{2p}+R_2')}$ conditionally independent codewords $v_2^n(m_1,m_2,j,j')$ with each symbol i.i.d according to $p_{V_2|U_1,Q}$, $j\in\{1,...,2^{nR_{2p}}\}$, $ j'\in\{1,...,2^{nR_2'}\}$. As such, the rate of receiver 1's message, denoted by $R_1$, is given by $R_1=R_{1c}+R_{1p}$. Similarly the rate of receiver 2, denoted by $R_2$, is given by $R_2=R_{2c}+R_{2p}$.

Receiver 3's message is split into common message $k$, to be decoded by both receiver 2 and receiver 3 and private message $l$, to be decoded only by receiver 3. We generate $2^{nT_3}$ independent codewords $u_2^n(k)$ with each symbol i.i.d according to $p_{U_2|Q}$, $k\in\{1,...,2^{nT_3}\}$. For each $u_2^n(k)$, we generate $2^{nS_3}$ conditionally independent codewords $x_2^n(k,l)$ with each symbol i.i.d according to $p_{X_2|U_2,Q}$, $l\in\{1,...,2^{nS_3}\}$. Hence the rate of receiver 3, denoted by $R_3$, is given by $R_3=T_3+S_3$.

\textit{Encoding}:

Given message quadruple  $(m_1,i,m_2,j)$, transmitter 1 finds a pair $(i',j')$ such that
\begin{align*}
    (v_1^n(m_1,m_2,i,i'),v_2^n(m_1,m_2,j,j'))\in A_{\epsilon}^{(n)}(V_1,V_2).
\end{align*}
If there is one or more such pairs, the transmitter chooses one and generates $x_1^n$ as
\begin{align*}
  x_{1,k}=f(v_{1,k}(m_1,m_2,i,i'),v_{2,k}(m_1,m_2,j,j')),\quad k\in\{1,...,n\}
\end{align*}
where $f(\cdot)$ is some function. If there is no such pair, an error is declared and a predefined codeword is sent. Transmitter 2 sends codeword $x_2^n(k,l)$ for message pair $(k,l)$. Without loss of generality, in the following we assume $(m_1,i,m_2,j,k,l)=(1,1,1,1,1,1)$ is sent.

\textit{Decoding}:

Receiver 1 looks for $(\hat{m}_1,\hat{m}_2,\hat{i},\hat{i}')$ such that
\begin{align*}
(q^n, u_1^n(\hat{m}_1,\hat{m}_2) ,v_1^n(\hat{m}_1,\hat{m}_2,\hat{i},\hat{i}'), y_1^n)\in A_{\epsilon}^{(n)}(Q,U_1,V_1,Y_1).
\end{align*}
If there is no such quadruple or some such quadruple with either $\hat{m}_1\neq 1$ or $\hat{i}\neq 1$ or both, there is an error. Receiver 2 looks for $(\hat{m}_1,\hat{m}_2,\hat{j},\hat{j}',\hat{k})$ such that
\begin{align*}
    (q^n,u_1^n(\hat{m}_1,\hat{m}_2), v_2^n(\hat{m}_1,\hat{m}_2,\hat{j},\hat{j}'), u_2^n(\hat{k}),y_2^n)\in A_{\epsilon}^{(n)}(Q,U_1,V_2,U_2,Y_2).
\end{align*}
If there is no such quintuple or some such quintuple with either $\hat{m}_2\neq 1$ or $\hat{j}\neq 1$ or both, there is an error. Receiver 3 looks for unique $(\hat{k},\hat{l})$ such that
\begin{align*}
    (q^n,u_2^n(\hat{k}),x_2^n(\hat{k},\hat{l}),y_3^n)\in A_{\epsilon}^{(n)}(Q,U_2,X_2,Y_3).
\end{align*}
If there is none or more than one such pair, there is an error.

Note that with these decoding criteria, erroneous decoding of non-intended messages does not constitute error events, which could potentially enlarge the achievable rate region compared to the case where all messages related to the joint-typicality decoding are required to be successfully decoded. For example, receiver 1 is only interested in message $m_1$ and $i$. Hence for $\hat{m}_1=1,\hat{i}=1$ and some $\hat{m}_2\neq 1$, event
\begin{align*}
  (q^n,u_1^n(\hat{m}_1,\hat{m}_2),v_1^n(\hat{m}_1,\hat{m}_2,\hat{i},\hat{i}'), y_1^n)\in A_{\epsilon}^{(n)}(Q,U_1,V_1,Y_1)
\end{align*}
does not cause an error even though $\hat{m}_2\neq m_2$. Similarly for $Y_2$, messages $m_1$ and $k$ are irrelevant. This has been previously leveraged in \cite{CMG} to obtain a compact Han-Kobayashi region for the interference channel.

\textit{Analysis of error probability}:

Given $(m_1,m_2,i,j)$, with high probability there is at least one $(i',j')$ pair such that $(v_1^n(m_1,m_2,i,i'),$ $v_2^n(m_1,m_2,j,j'))$ is jointly typical if $R_1'+R_2'>I(V_1;V_2|U_1,Q)$ due to mutual covering lemma \cite{El Gamal}.

\textit{At receiver 1}:
Using standard techniques from \cite{El Gamal}, where all error events are first determined using a joint pmf factorization table and then analyzed individually using packing lemma \cite{El Gamal}, it can be shown that the error probability at receiver 1 can be made arbitrarily small if
\begin{align}
    R_{1p}+R_1'&\leq I(V_1;Y_1|U_1,Q) \label{cnstinput1}\\
    R_{1c}+R_{2c}+R_{1p}+R_1'&\leq I(V_1;Y_1|Q).\notag
\end{align}

\textit{At receiver 2}:
Similarly it can be shown that the error probability at receiver 2 can be made arbitrarily small if
\begin{align}
    R_{2p}+R_2'&\leq I(V_2;Y_2|U_1,U_2,Q) \label{cnstinput2}\\
    R_{2p}+R_2'+T_3&\leq I(V_2,U_2;Y_2|U_1,Q) \notag\\
    R_{1c}+R_{2c}+R_{2p}+R_2' &\leq I(V_2;Y_2|U_2,Q) \notag\\
    R_{1c}+R_{2c}+R_{2p}+R_2'+T_3 &\leq I(V_2,U_2;Y_2|Q).\notag
\end{align}

\textit{At receiver 3}:
Similarly it can be shown that the error probability at receiver 3 can be made arbitrarily small if
\begin{align*}
    S_3&\leq I(X_2;Y_3|U_2,Q) \\
    T_3+S_3&\leq I(X_2;Y_3|Q).
\end{align*}

After collecting all inequalities, we apply Fourier-Motzkin elimination with $R_1=R_{1c}+R_{1p}$, $R_2=R_{2c}+R_{2p}$ and $R_3=T_3+S_3$ to derive a compact expression of the rate region, i.e. $\hat{\mathcal{R}}_P$. This step is lengthy but standard, which we omit for conciseness.
\end{IEEEproof}

%------------------------------------------------------------------------------------------------------------------
\section{Proof of Corollary \ref{CoroR2}}
\label{ProofCoroR2}
\begin{IEEEproof}
Let $\mathcal{R}'=\bigcup_{P\in\mathcal{P}}\hat{\mathcal{R}}_P'$, where $\hat{\mathcal{R}}_P'$ is the same as $\hat{\mathcal{R}}_P$ given in Lemma \ref{R3hat} except that (\ref{rdt2}) is removed and $\mathcal{P}$ is the set of joint input distributions that factor as in (\ref{pdffactor}). Since (\ref{rdt2}) is redundant by Proposition \ref{equivalence}, we have $\mathcal{R}' = \mathcal{R}$. Now we fix $Q=\phi$ and evaluate $\hat{\mathcal{R}}_P'$ with $V_1=X_1$, $V_2=U_1$ to obtain a region specified by the same inequalities defining $\mathcal{R}_{2,P}$ plus one extra inequality
\begin{align*}
    R_3\leq I(U_2;Y_2|U_1)+I(X_2;Y_3|U_2).
\end{align*}
Using the same argument for Proposition \ref{equivalence}, we can show that the above inequality is redundant. Hence $\mathcal{R}_{2,P}$ is achievable. By taking the union of $\mathcal{R}_{2,P}$ for all $P\in\mathcal{P}'$, we obtain $\mathcal{R}_2$.
\end{IEEEproof}

%------------------------------------------------------------------------------------------------------------------
\section{Proof of Proposition \ref{convexification}}
\label{ProofPropositionConv}
\begin{IEEEproof}
We prove for $\mathcal{R}_1$ and the case of $\mathcal{R}_2$ follows similarly. Without loss of generality, we let $Q$ take two values 1, 2 with probability $\alpha$ and $\bar{\alpha}=1-\alpha$, $0\leq\alpha\leq1$. We consider two tuples $(U_1^i,X_1^i,U_2^i,X_2^i,Y_1^i,Y_2^i)$ where $i=1,2$. For $Q=i$, define $U_{1,Q}=U_1^i$, $U_{2,Q}=U_2^i$, $X_1=X_1^i$, $X_2=X_2^i$, $Y_1=Y_1^i$ and $Y_2=Y_2^i$. Then we have Markov chain $(Q,U_{1,Q},U_{2,Q})\rightarrow (X_1,X_2)\rightarrow(Y_1,Y_2)$.

For (\ref{ts1}), we have
\begin{align*}
    \alpha I(U_1^1;Y_1^1) + \bar{\alpha} I(U_1^2;Y_1^2)=I(U_{1,Q};Y_2|Q)\leq I(U_{1,Q},Q;Y_2).
\end{align*}

For (\ref{ts2}), we have
\begin{align*}
    \alpha I(X_2^1;Y_3^1) + \bar{\alpha} I(X_2^2;Y_3^2)&=I(X_2;Y_3|U_{2,Q},Q) +  I(U_{2,Q};Y_3|Q)\\
    &\leq  I(X_2;Y_3|U_{2,Q},Q) +  I(U_{2,Q},Q;Y_3)=I(X_2;Y_3).
\end{align*}
Similarly, we can show that the convex combinations of the right-hand sides of (\ref{largestR22}), (\ref{largestR33}) are respectively less or equal to
\begin{align*}
    &I(U_{1,Q},Q;Y_1)+I(X_1;Y_2|U_{1,Q},U_{2,Q},Q),\\
    &I(U_{1,Q},Q;Y_1)+I(X_1,U_{2,Q},Q;Y_2|U_{1,Q},Q)+I(X_2;Y_2|U_{2,Q},Q).
\end{align*}
Redefining $U_1=(U_{1,Q},Q)$ and $U_2=(U_{2,Q},Q)$, we see that the time-sharing region is always contained in $\mathcal{R}_1$.
\end{IEEEproof}

%---------------------------------------------------------------------------------------------------------------
\section{Proof of Lemma \ref{DEXPR2minus}}
\label{proofDEXPR2minus}
\begin{IEEEproof}
To simplify the notation, let $\Omega$ denote the set of all DExPs of $\mathcal{R}_{2,P}$. For some constant rate $R_1'$, let $\mathcal{R}_{2,P}(R_1')$ denote the region $\mathcal{R}_{2,P}$ with $R_1=R_1'$, i.e. $\mathcal{R}_{2,P}(R_1') =\{(R_2,R_3): (R_1',R_2,R_3) \in\mathcal{R}_{2,P}\}$. Then the set of all DExPs of $\mathcal{R}_{2,P}(R_1')$ is denoted by $\Omega(R_1')$. Similarly, we could also define $\mathcal{R}_{2,P}(R_i')$ and $\Omega(R_i')$ for $i=2,3$, and $\mathcal{R}_{2,P}(R_k',R_l')$ and $\Omega(R_k',R_l')$ for $k,l=1,2,3$ and $k< l$.

The region $\mathcal{R}_{2,P}$ is given by a system of linear inequalities. Since DExPs are ExPs by definition, which can be found by solving the system of linear equations given by some active constraints, one approach to determine $\Omega$ is to consider all possible combinations of active constraints whose corresponding system of linear equations admits a unique solution and then compare the obtained ExPs one by one. There are totally eight inequalities defining $\mathcal{R}_{2,P}$ (including $R_i\geq 0$, $i=1,2,3$), making this approach tedious. Fortunately, some properties of DExPs can be used to simply the procedure and make it systematic so that we won't overlook any DExP.

Let $R_i^*$ denote the largest admissible $R_i$ in $\mathcal{R}_{2,P}$. Then DExPs can be sorted into four categories:
\begin{enumerate}
\item Case 1: $(R_1^*,R_2,R_3)\in\Omega$ for some $R_i\leq R_i^*$, $i=2,3$
\item Case 2: $(R_1,R_2^*,R_3)\in\Omega$ for some $R_i\leq R_i^*$, $i=1,3$
\item Case 3: $(R_1,R_2,R_3^*)\in\Omega$ for some $R_i\leq R_i^*$, $i=1,2$
\item Case 4: $(R_1,R_2,R_3)\in\Omega$ for some $R_i<R_i^*$, $i=1,2,3$
\end{enumerate}

Note that Case 1, 2, 3 are not mutually exclusive. The rationale of such division is, by considering Case 1, 2, 3, a higher dimensional ($n=3$) problem can be reduced to a lower one ($n=1$ or $n=2$) and for the irreducible Case 4, the additional constraints $R_i<R_i^*$ will simplify the problem. This point will be made clear as we proceed in the following.

\textit{Case 1:}

The largest admissible $R_1^*=I(X_1;Y_1|U_1)+I(U_1;Y_2|U_2)$ is obtained by setting $R_2=0$ in (\ref{largestR1}). Fixing $R_1'=R_1^*$, $R_2'=0$, we have $(R_1',R_2',R_3)\in\Omega$ iff $R_3\in\Omega(R_1',R_2')$. Since $\sup_{R_3\in\mathcal{R}_{2,P}(R_1',R_2')} R_3 =\min\{ I(X_2;Y_3) , I(U_2;Y_2)+I(X_2;Y_3|U_2) \}$, we obtain DExP $D$.

\textit{Case 2:}

The largest admissible $R_2^*=I(U_1;Y_2|U_2)$ is given by (\ref{largestR2}). Fixing $R_2'=R_2^*$, we have $(R_1,R_2',R_3) \in \Omega$ iff $(R_1,R_3)\in\Omega(R_2')$ and
\begin{align*}
    \mathcal{R}_{2,P}(R_2')=
    \left\{\begin{aligned}
    &(R_1,R_3):  R_1,R_3\geq 0,\ R_1 \leq I(X_1;Y_1|U_1) \triangleq r_a ,   \\
    &R_3\leq \min\{I(X_2;Y_3),\ I(U_2;Y_2)+I(X_2;Y_3|U_2)\}  \triangleq r_b
    \end{aligned}\right\}.
\end{align*}
It is easy to see that $\Omega(R_2')=\{(r_a, r_b)\}$, yielding DExP $A$.

\textit{Case 3:}

The largest admissible $R_3$ is given by $R_3^*=\min\{I(X_2;Y_3),\ I(U_1,U_2;Y_2)+I(X_2;Y_3|U_2)\} $. If $R_3^*=I(X_2;Y_3)$, fixing $R_3'=R_3^*$ and we have
\begin{align*}
    \mathcal{R}_{2,P}(R_3')=
    \left\{\begin{aligned}
    &(R_1,R_2):  R_1,R_2\geq 0,\ R_2\leq \min\{I(U_1;Y_2|U_2),\ I(U_1,U_2;Y_2)-I(U_2;Y_3)\}\triangleq r_c,  \\
    &R_1+R_2 \leq I(X_1;Y_1|U_1)+ \min\{I(U_1;Y_2|U_2), I(U_1,U_2;Y_2)-I(U_2;Y_3)\} \triangleq r_d  \\
    \end{aligned}\right\}.
\end{align*}
Note that $I(U_1,U_2;Y_2)-I(U_2;Y_3)\geq 0$ in this case. It is easy to see $\Omega(R_3')=\{(r_d-r_c,r_c),(r_d,0)\}$, resulting in two DExPs,
\begin{align*}
    B'&=(\ I(X_1;Y_1|U_1),\ \min\{I(U_1;Y_2|U_2),\ I(U_1,U_2;Y_2)- I(U_2;Y_3)\},\ I(X_2;Y_3) \ ),\\
    C'&=(\ I(X_1;Y_1|U_1)+ \min\{I(U_1;Y_2|U_2),\ I(U_1,U_2;Y_2)-I(U_2;Y_3)\},\ 0, \  \ I(X_2;Y_3)\ ).
\end{align*}
If $R_3^*=I(U_1,U_2;Y_2)+I(X_2;Y_3|U_2)$, which is given by (\ref{largestR3}) by setting $R_2=0$, fixing $R_2'=0$, $R_3'=R_3^*$ and we obtain $\mathcal{R}_{2,P}(R_2',R_3')=\{ R_1: 0\leq R_1\leq  I(X_1;Y_1|U_1)  \}$. Hence we obtain one DExP
\begin{align*}
    E'=\big( I(X_1;Y_1|U_1),\ 0,\ I(U_1,U_2;Y_2)+I(X_2;Y_3|U_2)  \big).
\end{align*}
Combing the two cases, we rewrite $B'$ and $E'$ collectively as $B$, and $C'$ and $E'$ collectively as $C$.

\textit{Case 4:}

Under the condition $R_i<R_i^*$, $i=1,2,3$, $\mathcal{R}_{2,P}$ is given by
\begin{align}
    R_1&<I(X_1;Y_1|U_1)+I(U_1;Y_2|U_2)  \label{vioR1}\\
    R_2&< I(U_1;Y_2|U_2)\label{vioR2}\\
    R_3&< \min\{ I(X_2;Y_3),\ I(U_1,U_2;Y_2)+ I(X_2;Y_3|U_2) \} \label{vioR3}\\
    R_1+R_2&\leq I(X_1;Y_1|U_1)+I(U_1;Y_2|U_2)\label{constraint1}\\
    R_2+R_3&\leq I(U_1,U_2;Y_2)+I(X_2;Y_3|U_2)\label{constraint2}\\
    R_1+R_2+R_3&\leq I(X_1;Y_1|U_1)+I(U_1,U_2;Y_2)+ I(X_2;Y_3|U_2)\label{constraint3}\\
    R_1,R_2,R_3&\geq 0.  \label{constraint4}
\end{align}
As mentioned before, DExPs are ExPs and hence the solutions of systems of linear equations given by some active constraints. Next we will first consider all possible combinations of active constraints defining dominant faces, i.e. (\ref{constraint1})-(\ref{constraint3}) and then add additional active constraints from (\ref{constraint4}) as needed to ensure the resulting system has a unique solution.

If (\ref{constraint1})-(\ref{constraint3}) are all active, from (\ref{constraint2}), (\ref{constraint3}) we get $R_1=I(X_1;Y_1|U_1)$ and further with (\ref{constraint1}), we get $R_2=I(U_1;Y_2|U_2)$, which violates (\ref{vioR2}). If only (\ref{constraint1}) and (\ref{constraint2}) are active, since the corresponding system of linear equations does not have a unique solution, we choose one additional active constraint from (\ref{constraint4}). However, the obtained solution violates one of (\ref{vioR1})-(\ref{vioR3}). We can proceed similarly for the remaining six possible combinations and none of them produces a DExP. Overall we conclude that there is no DExP in Case 4.
\end{IEEEproof}

%------------------------------------------------------------------------------------------------------------------
\section{Proof of Lemma \ref{DEXPRpprimeminus}}
\label{proofDEXPRpprimeminus}
\begin{IEEEproof}
We use similar notations from Appendix \ref{proofDEXPR2minus}, which are now defined over $\mathcal{R}_{(2),P}$ instead of $\mathcal{R}_{2,P}$. As we can see, $R_1$ is disassociated with $R_2$, $R_3$. Hence the DExPs of $\mathcal{R}_{(2),P}$ are of the form $(R_1^*,R_2,R_3)$, where $R_1^*=I(X_1;Y_1|U_1)$. Fixing $R_1'=R_1^*$, we have
\begin{align*}
    \mathcal{R}_{(2),P}(R_1')=
    \left\{\begin{aligned}
    (R_2,R_3): &R_2,R_3\geq 0,\ R_2\leq I(U_1;Y_2|U_2)\triangleq r_e,\ R_3\leq I(X_2;Y_3) \triangleq r_f, \\
    &R_2+R_3\leq I(U_1,U_2;Y_2)+I(X_2;Y_3|U_2) \triangleq r_g \\
    \end{aligned}\right\}.
\end{align*}

If $r_g\geq r_f$, i.e. $I(U_1,U_2;Y_2)\geq I(U_2;Y_3)$, $\mathcal{R}_{(2),P}(R_1')$ is depicted in Fig. \ref{Region}(a) and $\Omega(R_1')=\{T_1,T_2\}=\{(r_e,\min\{r_f,r_g-r_e\}),(\min\{r_e,r_g-r_f\},r_f)\}$, yielding two DExPs, $A$ and
\begin{align*}
    B'&=(\ I(X_1;Y_1|U_1),\ \min\{I(U_1;Y_2|U_2),\ I(U_1,U_2;Y_2)- I(U_2;Y_3) \} ,\  I(X_2;Y_3)\ ).
\end{align*}

If $r_g< r_f$, i.e. $I(U_1,U_2;Y_2)< I(U_2;Y_3)$, $\mathcal{R}_{(2),P}(R_1')$ is depicted in Fig. \ref{Region}(b) and $\Omega(R_1')=\{T_1,T_3\}=\{(r_e,\min\{r_f,r_g-r_e\}),(0,r_g)\}$, yielding one more DExP of $\mathcal{R}_{(2),P}$
\begin{align*}
    C'&=(\ I(X_1;Y_1|U_1),\ 0 ,\ I(U_1,U_2;Y_2)+I(X_2;Y_3|U_2)\ ).
\end{align*}

Note that $B'$ and $C'$ can be rewritten collectively as $B$.
\begin{figure}
  \centering
  \subfloat[]{\includegraphics[width=0.18\textwidth]{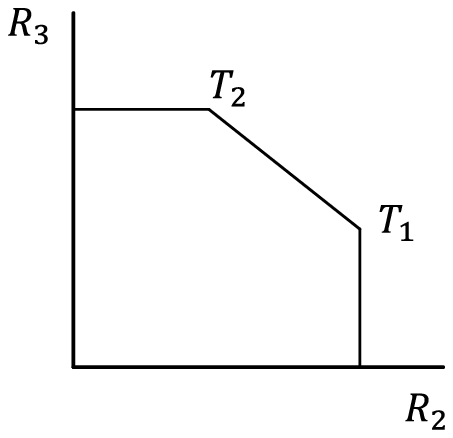}}
  \qquad
  \subfloat[]{\includegraphics[width=0.18\textwidth]{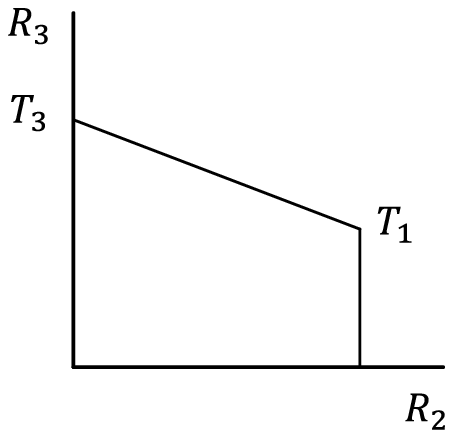}}
  \caption{Shape of $\mathcal{R}_{(2),P}(R_1')$ in the proof of Lemma \ref{DEXPRpprimeminus}}
  \label{Region}
\end{figure}
\end{IEEEproof}

%------------------------------------------------------------------------------------------------------------------
\section{Proof of Lemma \ref{DEXPR1minus}}
\label{proofDEXPR1minus}
\begin{IEEEproof}
The proof is similar to that of Lemma \ref{DEXPR2minus} and we use similar notations from Appendix \ref{proofDEXPR2minus}, which are now defined over $\mathcal{R}_{1,P}$ instead of $\mathcal{R}_{2,P}$. Since DExPs can be sorted into four categories as in Appendix \ref{proofDEXPR2minus}, we next discuss case by case.

\textit{Case 1:}

The largest admissible $R_1^*=I(U_1;Y_1)$. Fixing $R_1'=R_1^*$, we have $(R_1',R_2,R_3) \in \Omega$ iff $(R_2,R_3) \in \Omega(R_1')$ and
\begin{align*}
    \mathcal{R}_{1,P}(R_1')=
    \left\{\begin{aligned}
    (R_2,R_3): & R_2,R_3\geq 0,\ R_2\leq I(X_1;Y_2|U_1,U_2),\ R_3\leq I(X_2;Y_3),  \\
    &R_2+R_3\leq I(X_1,U_2;Y_2|U_1)+I(X_2;Y_3|U_2)
    \end{aligned}\right\}.
\end{align*}
Similar to Appendix \ref{proofDEXPRpprimeminus}, we can show that there are two DExPs, $E$ and $F$.

\textit{Case 2:}

The largest admissible $R_2^*=I(U_1;Y_1)+I(X_1;Y_2|U_1,U_2)$ is obtained by setting $R_1=0$ in (\ref{largestR22}). Fixing $R_1'=0$ and $R_2'=R_2^*$, we have $\mathcal{R}_{1,P}(R_1',R_2')=\{R_3:0\leq R_3\leq \min\{ I(X_2;Y_3) ,\ I(U_2;Y_2|U_1)+I(X_2;Y_3|U_2) \}\}$, resulting in DExP $G$.

\textit{Case 3:}

The largest admissible $R_3^*=\min\{ I(X_2;Y_3),\ I(U_1;Y_1)+I(X_1,U_2;Y_2|U_1)+I(X_2;Y_3|U_2)\}$.

{\em 1.} If $I(U_2;Y_3)\leq I(U_1;Y_1)+I(X_1,U_2;Y_2|U_1)$, $R_3^*=I(X_2;Y_3)$. Fixing $R_3'=R_3^*$, we have
\begin{align*}
    \mathcal{R}_{1,P}(R_3')=
    \left\{\begin{aligned} &(R_1,R_2):R_1,R_2\geq 0,\  R_1\leq I(U_1;Y_1) \triangleq r_i,\\
    &R_1+R_2 \leq I(U_1;Y_1)+ I(X_1;Y_2|U_1,U_2)+ \min\{0,\ I(U_2;Y_2|U_1)-I(U_2;Y_3)\} \triangleq r_j\\
    \end{aligned}\right\}.
\end{align*}
Note that $r_j\geq 0$ in this case. This case is similar to that in Appendix \ref{proofDEXPRpprimeminus}, and we can show
\begin{enumerate}
\item If $I(U_2;Y_3)\leq I(X_1,U_2;Y_2|U_1)$, $\Omega(R_3')=\{(0,r_j),(r_i,r_j-r_i)\}$.
\item If $I(X_1,U_2;Y_2|U_1)<I(U_2;Y_3)\leq I(U_1;Y_1)+I(X_1,U_2;Y_2|U_1)$, $\Omega(R_3')=\{(0,r_j),(r_j,0)\}$.
\end{enumerate}
Finally, we collectively write the obtained DExPs as
\begin{align*}
    H =(\ 0,\ r_j,\ R_3^*\ ),\ I = (\ \min\{r_i,r_j\},\ [r_j-r_i]^+,\ R_3^* \ ).
\end{align*}

{\em 2.} If $I(U_2;Y_3)> I(U_1;Y_1)+I(X_1,U_2;Y_2|U_1)$, $R_3^*=I(U_1;Y_1)+I(X_1,U_2;Y_2|U_1)+I(X_2;Y_3|U_2)$, which is obtained by setting $R_1=R_2=0$ in (\ref{largestR33}). In this case we find one DExP $J$.

\textit{Case 4:} Similar to that in Appendix \ref{proofDEXPR2minus}, it can be shown that there is no DExP in this case.
\end{IEEEproof}

%---------------------------------------------------------------------------------------------------------------
\section{Proof of Theorem \ref{capastrong}}
\label{proofcapastrong}
For the converse proof, we will use a technique proposed in \cite{Nair}. Specifically, we need the following lemma that can be easily proved using similar arguments as in \cite[Lemma 1]{Nair}.

\begin{lemma}
\label{lemmalessnoisy}
In a DM-BIC with $Y_1\prec_o Y_2$, if $W\rightarrow(X_1^n,X_2^n)\rightarrow(Y_1^n,Y_2^n)$ form a Markov chain, then the following holds:
\begin{align*}
    I(Y_2^{i-1};Y_{2,i}|W)\geq I(Y_1^{i-1};Y_{2,i}|W),\quad 1\leq i \leq n.
\end{align*}
\end{lemma}

\begin{IEEEproof}[Proof of Theorem \ref{capastrong}]
Specializing $\mathcal{R}_{(1),P}$ in Proposition \ref{Rprime} with $U_2=X_2$, the region given in Theorem \ref{capastrong} is achievable. For the converse, we define $U_i=(W_1,Y_1^{i-1})$. For some $\epsilon_n$ such that $\lim_{n\rightarrow\infty}\epsilon_n=0$, by Fano's inequality we have
\begin{align*}
    n(R_1-\epsilon_n)&\leq I(W_1;Y_1^n)=\sum_{i=1}^n I(W_1;Y_{1,i}|Y_1^{i-1}) \leq \sum_{i=1}^n I(U_i;Y_{1,i}).
\end{align*}

To bound $R_2$, we proceed as follows
\begin{align}
    n(R_2-\epsilon_n)&\leq I(W_2;Y_2^n|W_1,X_2^n)\notag\\
    &\leq \sum_{i=1}^n I(X_{1,i};Y_{2,i}|W_1,X_{2,i},Y_2^{i-1})\notag\\
    &=\sum_{i=1}^n I(X_{1,i};Y_{2,i}|W_1,X_{2,i})-I(Y_2^{i-1};Y_{2,i}|W_1,X_{2,i})\notag\\
    &\leq \sum_{i=1}^n I(X_{1,i};Y_{2,i}|W_1,X_{2,i})-I(Y_1^{i-1};Y_{2,i}|W_1,X_{2,i}) \label{converse1}\\
    &=\sum_{i=1}^n I(X_{1,i};Y_{2,i}|U_i,X_{2,i}),\notag
\end{align}
where (\ref{converse1}) follows from Lemma \ref{lemmalessnoisy}.

Now we consider $R_2+R_3$. The strong interference condition implies $I(X_2^n;Y_2^n|X_1^n)\geq I(X_2^n;Y_3^n)$, \cite{Costa}.
\begin{align}
    n(R_2+R_3-\epsilon_n)&\leq I(W_2;Y_2^n)+I(W_3;Y_3^n)\notag\\
    &\leq I(X_1^n;Y_2^n|W_1) + I(X_2^n;Y_2^n|X_1^n) \label{converse2}\\
    &=\sum_{i=1}^n I(X_{1,i},X_{2,i};Y_{2,i}|W_1,Y_2^{i-1})\notag\\
    &=\sum_{i=1}^n I(X_{1,i},X_{2,i};Y_{2,i}|W_1)- I(Y_2^{i-1};Y_{2,i}|W_1)\notag\\
    &\leq \sum_{i=1}^n I(X_{1,i},X_{2,i};Y_{2,i}|W_1)- I(Y_1^{i-1};Y_{2,i}|W_1) \label{converse3}\\
    &=\sum_{i=1}^n I(X_{1,i},X_{2,i};Y_{2,i}|U_{i}),\notag
\end{align}
where (\ref{converse2}) is due to data processing inequality and the strong interference condition, and (\ref{converse3}) follows from Lemma \ref{lemmalessnoisy}.

Finally, we have $n(R_3-\epsilon_n)\leq \sum_{i=1}^n I(X_{2,i};Y_{3,i})$. The proof is complete by redefining $U=(U_Q,Q)$, $X_{j,i}=X_{j}$ for $j=1,2$, and $Y_{l,i}=Y_{l}$, for $l=1,2,3$, where $Q$ is a uniformly distributed random variable on $\{1,...,n\}$.
\end{IEEEproof}

%---------------------------------------------------------------------------------------------------------------
\section{Proof of Theorem \ref{theo:outerbound2}}
\label{prooftheo:outerbound2}
\begin{IEEEproof}
The bound $R_3\leq \mathcal{C}(P_2)$ is straightforward. To bound $R_1$, we have
\begin{align*}
 n(R_1-\epsilon_n) &\leq I(W_1;Y_1^n)=h(Y_1^n)-h(Y_1^n|W_1)\leq \frac{n}{2}\log[2\pi e(1+P_1)]-h(Y_1^n|W_1).
\end{align*}
Since $\frac{n}{2}\log2\pi e=h(Y_1^n|X_1^n)\leq h(Y_1^n|W_1)\leq h(Y_1^n)=\frac{n}{2}\log[2\pi e(1+P_1)]$, we must have $h(Y_1^n|W_1)=\frac{n}{2}\log[2\pi e(1+\alpha P_1)]$ for some $\alpha\in[0,1]$. Therefore we obtain bound (\ref{EQ3}) on $R_1$.

Before bounding $R_2$, let us first consider the term $h(\sqrt{b}X_2^n+Z_2^n)$.

\emph{Case I}: $b<1$\\
In this case, we have
\begin{align}
    h(\sqrt{b}X_2^n+Z_2^n)&=h(\sqrt{b}(X_2^n+Z_3^n)+\sqrt{1-b}\tilde{Z}^n) \label{BR2EQ2}\\
    %&\geq \frac{n}{2}\log\left[ 2^{\frac{2}{n}h(\sqrt{b}(X_2^n+Z_3^n))} + 2^{\frac{2}{n}h(\sqrt{1-b}\tilde{Z}^n)} \right] \label{BR2EQ3}\\
    &\geq\frac{n}{2}\log\left[ 2^{\frac{2}{n}h(\sqrt{b}(X_2^n+Z_3^n))} + 2\pi e(1-b) \right]\label{BR2EQ4}
\end{align}
where in (\ref{BR2EQ2}), $\tilde{Z}\sim\mathcal{N}(0,1)$ is independent of $Z_3$, and (\ref{BR2EQ4}) is due to EPI \cite{Cover}. Since
\begin{align*}
    n(R_3-\epsilon_n)\leq I(X_2^n;Y_3^n)=h(X_2^n+Z_3^n)-\frac{n}{2}\log(2\pi e),
\end{align*}
we have $h(\sqrt{b}(X_2^n+Z_3^n))\geq nR_3+\frac{n}{2}\log(2\pi eb)$ ($\epsilon_n$ is removed for conciseness). Continuing from (\ref{BR2EQ4}), we have
\begin{align*}
    h(\sqrt{b}X_2^n+Z_2^n)&\geq \frac{n}{2}\log\left[ 2\pi eb 2^{2R_3}+2\pi e(1-b)\right]\\
    &=\frac{n}{2}\log\left[1+b(2^{2R_3}-1)\right]+\frac{n}{2}\log 2\pi e=n\xi(b)+\frac{n}{2}\log 2\pi e .
\end{align*}

\emph{Case II}: $b\geq 1$\\
In this case, we have
\begin{align}
    h(\sqrt{b}X_2^n+Z_2^n)&=h(X_2^n+ \tfrac{1}{\sqrt{b}}Z_2^n)+\frac{n}{2}\log b \notag \\
    &\geq \frac{n}{2}\log\left[(1-\tfrac{1}{b})2^{\frac{2}{n}h(X_2^n)}+\tfrac{1}{b}2^{\frac{2}{n}h(X_2^n+Z_2^n)}\right] +\frac{n}{2}\log b \label{BR2EQ5}\\
    &\geq \frac{n}{2}\log\left[2^{\frac{2}{n}h(X_2^n+Z_2^n)}\right]\notag\\
    &\geq nR_3+\frac{n}{2}\log 2\pi e=n\xi(b)+\frac{n}{2}\log 2\pi e \label{BR2EQ6}
\end{align}
where (\ref{BR2EQ5}) is due to Costa's EPI \cite{NEPI} and (\ref{BR2EQ6}) is due to $h(X_2^n+Z_3^n)\geq nR_3+\frac{n}{2}\log 2\pi e$.

To summarize, we have proved that $h(\sqrt{b}X_2^n+Z_2^n)\geq \xi(b)$ where $\xi$ is defined in (\ref{xi}). Before we bound $R_2$, we need to first bound $h(Y_2^n|W_1)$:
\begin{align}
 h(Y_2^n|W_1)&= h(X_1^n+\tfrac{1}{\sqrt{a}}(\sqrt{b}X_2^n+Z_2^n)|W_1)+\frac{n}{2}\log a \notag\\
&\leq h(X_1^n+\tilde{Z}^n|W_1)+\frac{n}{2}\log a + 0.5n, \label{OBEQ1}
\end{align}
where $\tilde{Z}\sim \mathcal{N}(0,\frac{1+bP_2}{a})$. To obtain (\ref{OBEQ1}), we use the fact that for an arbitrarily distributed additive noise channel, Gaussian input incurs no more than 0.5 bits loss compared to the optimal input distribution \cite{Zamir}, which in this context translates to $h(X_1^n+\tfrac{1}{\sqrt{a}}(\sqrt{b}X_2^n+Z_2^n)|W_1) \leq h(X_1^n+\tfrac{1}{\sqrt{a}}(\sqrt{b}X_{2,G}^n+Z_2^n)|W_1)+0.5n$ with $X_{2,G}\sim\mathcal{N}(0,P_2)$. Since $a\geq 1+bP_2$, using EPI we have
\begin{align*}
 h(Y_1^n|W_1)&=h(X_1^n+Z_1^n|W_1)\geq\frac{n}{2}\log\left[ 2^{\frac{2}{n}h(X_1^n+\tilde{Z}^n|W_1)} + 2\pi e(1-\tfrac{1+bP_2}{a}) \right].
\end{align*}
Therefore $h(X_1^n+\tilde{Z}^n|W_1)\leq \frac{n}{2}\log\left[ 2\pi e(\alpha P_1+\frac{1+bP_2}{a}) \right]$. Continuing from (\ref{OBEQ1}), we obtain
\begin{align}
 h(Y_2^n|W_1)\leq \frac{n}{2}\log(1+a\alpha P_1+bP_2 ) + \frac{n}{2}\log 2\pi e + 0.5n. \label{neweq1}
\end{align}

We are now in position to bound $R_2$. Due to Fano's inequality and data processing inequality, we have
\begin{align*}
    n(R_2-\epsilon_n) &\leq I(X_1^n;Y_2^n|W_1)\\
    &= h(Y_2^n|W_1) - h(\sqrt{b}X_2^n+Z_2^n) \\
    &\leq \frac{n}{2}\log(1+a\alpha P_1+bP_2 )-n\xi(b)+0.5n.
\end{align*}
Therefore we obtain (\ref{EQ}). Similarly, to obtain (\ref{EQ2}), we have
\begin{align}
    n(R_2-\epsilon_n)&\leq I(X_1^n;Y_2^n) \notag\\
    &= h(\sqrt{a}X_1^n+\sqrt{b}X_2^n+Z_2^n) - h(\sqrt{b}X_2^n+Z_2^n) \notag\\
    &\leq \frac{n}{2}\log(1+aP_1+bP_2) - n\xi(b),\notag
\end{align}
where the last inequality is due to the fact that Gaussian distribution maximizes entropy given a covariance constraint. Finally to obtain (\ref{EQ4}), we consider
\begin{align}
    n(R_2-\epsilon_n)&\leq I(X_1^n;Y_2^n|W_1,X_2^n) \notag \\
    &= h(X_1^n+\tfrac{1}{\sqrt{a}}Z_2^n|W_1) - h(Z_2^n) +\tfrac{n}{2}\log a \notag \\
    &\leq \tfrac{n}{2}\log[2\pi e(\tfrac{1}{a}+\alpha P_1)] - \tfrac{n}{2}\log(2\pi e) +\tfrac{n}{2}\log a \label{neweq} \\
    &=n\mathcal{C}(a\alpha P_1),\notag
\end{align}
where (\ref{neweq}) follows similar to (\ref{neweq1}).
\end{IEEEproof}

%---------------------------------------------------------------------------------------------------------------
\section{Proof of Theorem \ref{theo:halfbitsumcap}}
\label{prooftheo:halfbitsumcap}
\begin{IEEEproof}
For $b\geq 1+aP_1$, $R_s=C_s$ due to Corollary \ref{coroverystrong}. For $1\leq b< 1+aP_1$, the converse follows from the outer bound $\mathcal{O}_1$ given in Theorem \ref{theo:outerbound2}. One upper bound of the sum rate can be derived from (\ref{EQ3}) and (\ref{EQ}):
\begin{align*}
  R_1+R_2+R_3&\leq \max_{\alpha\in[0,1]}\left\{\mathcal{C}\left(\tfrac{\bar{\alpha}P_1}{1+\alpha P_1}\right) + \mathcal{C} (a\alpha P_1+bP_2) \right\}  + 0.5,\\
&=\max_{\alpha\in[0,1]}\frac{1}{2}\log\left( a(1+P_1) -\tfrac{a-1-bP_2}{1+\alpha P_1}(1+P_1) \right) + 0.5,\\
&=\mathcal{C}(aP_1+bP_2)+0.5.
\end{align*}
Also from (\ref{EQ3}), (\ref{EQ4}) and (\ref{EQ5}), we have
\begin{align*}
R_1+R_2+R_3&\leq \max_{\alpha\in[0,1]} \left\{\mathcal{C}\left(\tfrac{\bar{\alpha}P_1}{1+\alpha P_1}\right) + \mathcal{C}(a\alpha P_1)\right\} + \mathcal{C}(P_2) \\
&= \mathcal{C}(aP_1)+\mathcal{C}(P_2).
\end{align*}
Therefore, when $1\leq b< 1+aP_1$, $C_s\leq R_s+0.5$.

For $0\leq b< 1$, we use genie bounding approach \cite{Anna_Veer} to prove the converse. In the following, whenever we write $X_G$, the subscript G is used to indicate that the distribution is Gaussian. Consider the genie signal $S_3=X_2+\eta N_3$, where $N_3\sim\mathcal{N}(0,1)$ is correlated with $Z_3$ with correlation coefficient $\rho$ and $\eta$ is some constant. Now consider
\begin{align}
  n&(R_1+R_2+R_3 -\epsilon_n) \notag \\
&\leq I(W_1;Y_1^n)+I(X_1^n;Y_2^n|W_1)+I(X_2^n;Y_3^n,S_3^n) \notag \\
&=I(W_1;Y_1^n)+h(Y_2^n|W_1)-h(Y_2^n|X_1^n)+h(S_3^n)-h(S_3^n|X_2^n)+h(Y_3^n|S_3^n)-h(Y_3^n|S_3^n,X_2^n) \notag\\
&\leq I(W_1;Y_1^n)+h(Y_2^n|W_1)-h(Y_2^n|X_1^n)+h(S_3^n)-nh(S_{3,G}|X_{2,G})+nh(Y_{3,G}|S_{3,G})\notag \\
&\quad -nh(Y_{3,G}|S_{3,G},X_{2,G}) \label{Genie1},
\end{align}
where (\ref{Genie1}) is due to the fact $h(Y_3^n|S_3^n)\leq nh(Y_{3,G}|S_{3,G})$, a result of \cite[Lemma 1]{Anna_Veer}. Now we consider
\begin{align}
  h(S_3^n)-h(Y_2^n|X_1^n)&=h(X_2^n+\eta N_3^n)-h(X_2^n+\tfrac{1}{\sqrt{b}}Z_2^n)-\tfrac{n}{2}\log b\notag \\
&=h(X_2^n+\eta N_3^n)-h(X_2^n+\eta N_3^n+V^n)-\tfrac{n}{2}\log b\label{Genie2} \\
&=-I(V^n;X_2^n+\eta N_3^n+V^n)-\tfrac{n}{2}\log b\notag\\
&\leq -nI(V;X_{2,G}+\eta N_3+V)-\tfrac{n}{2}\log b\label{Genie3}
\end{align}
In (\ref{Genie2}) $V\sim\mathcal{N}(0,\frac{1}{b}-\eta^2)$ is independent of other random variables. Notice that $V$ exists only if $\frac{1}{b}\geq \eta^2$. Also (\ref{Genie3}) is due to the worst case noise result for an additive noise channel \cite{Diggavi}. Continuing from (\ref{Genie1}) we have
\begin{align*}
  n(R_1+R_2+R_3 -\epsilon_n) &\leq I(W_1;Y_1^n)+h(Y_2^n|W_1)-nh(Y_{2,G}|X_{1,G})+nI(X_{2,G};Y_{3,G},S_{3,G}).
\end{align*}
Due to \cite[Lemma 8]{Anna_Veer}, $I(X_{2,G};Y_{3,G},S_{3,G})=I(X_{2,G};Y_{3,G})$ iff $\eta\rho=1$. Therefore
\begin{align*}
  n(R_1+R_2+R_3 -\epsilon_n) &\leq I(W_1;Y_1^n)+h(Y_2^n|W_1)-nh(Y_{2,G}|X_{1,G})+nI(X_{2,G};Y_{3,G})
\end{align*}
under the condition: $b\leq \rho^2$ for some $\rho\in[0,1]$. There always exists such $\rho$ if $b<1$. From Appendix \ref{prooftheo:outerbound2}, we have $I(W_1;Y_1^n)\leq n\mathcal{C}\left(\tfrac{\bar{\alpha}P_1}{1+\alpha P_1}\right)$ and $h(Y_2^n|W_1)\leq \frac{n}{2}\log[2\pi e(1+a\alpha P_1+bP_2 )]+0.5n$. Therefore
\begin{align*}
 n(R_1+R_2+R_3 -\epsilon_n) &\leq \max_{\alpha\in[0,1]} \left\{ n\mathcal{C}\left(\tfrac{\bar{\alpha}P_1}{1+\alpha P_1}\right) + n\mathcal{C} (a\alpha P_1+bP_2)\right\} -n\mathcal{C}(bP_2) + n\mathcal{C}(P_2)  + 0.5n,\\
&=n\mathcal{C}(\tfrac{aP_1}{1+bP_2}) + n\mathcal{C}(P_2)+0.5n,
\end{align*}
i.e. $C_s\leq R_s+0.5$, when $0\leq b<1$.
\end{IEEEproof}

%---------------------------------------------------------------------------------------------------------------
\section{Proof of Corollary \ref{coro:s3}}
\label{proofcoro:s3}
\begin{IEEEproof}
We first specialize $\mathcal{R}_P$ in Theorem \ref{BinningRegion} with $V_1=U_1$ and $V_2=X_1$ to obtain a region $\mathcal{R}'_1$, which differs from $\mathcal{R}_1$, the inner bound given by Corollary \ref{CoroR1} for a DM-BIC with $Y_1\prec_o Y_2$, by having two extra inequalities
\begin{align*}
  R_1+R_2&\leq I(X_1;Y_2|U_2)\\
  R_1+R_2+R_3&\leq I(X_1,U_2;Y_2) + I(X_2;Y_3|U_2).
\end{align*}
Note that these inequalities are redundant if $Y_1\prec_o Y_2$. For some $\alpha_1\in[0,1]$, let $X_1=U_1+V'$, where $U_1\sim\mathcal{N}(0,\bar{\alpha}_1 P_1)$ and $V'\sim\mathcal{N}(0,\alpha_1 P_1)$ are independent. Similarly for some $\gamma_1\in[0,1]$ let $U_2=\sqrt{\gamma_1}X_{2c}$ and $X_2=\sqrt{\gamma_1}X_{2c}+\sqrt{\bar{\gamma}_1}X_{2p}$, where $X_{2c},X_{2p}\sim \mathcal{N}(0,P_2)$ and are independent. Evaluating $\mathcal{R}'_1$ with these random variables, we obtain an achievable rate region $\mathcal{S}_2$. Similarly, if we specialize $\mathcal{R}_P$ with $V_1=X_1$ and $V_2=U_1$ and evaluate the obtained region with Gaussian inputs similar to the above, we can obtain an achievable rate region $\mathcal{S}_3$. Finally we take convex hull of the two obtained rate regions.
\end{IEEEproof}

\end{document}